\documentclass[authoryear,12pt]{elsarticle}
\usepackage{graphicx}

\usepackage{amssymb,amsmath}
\usepackage{exscale}
\usepackage{makeidx,shortvrb,latexsym}
\usepackage{epstopdf}
\usepackage{tabularx, booktabs, multirow}
\usepackage{color}

\usepackage{float} 
\usepackage[toc,title,page]{appendix}
\usepackage{graphicx}
\DeclareGraphicsExtensions{.pdf,.png,.jpg}
\usepackage{wrapfig}
\usepackage{lscape}
\usepackage{rotating}
\usepackage{blindtext}
\usepackage{tikz}
\usepackage{caption}
\usepackage{subcaption}
\definecolor{chromeyellow}{rgb}{1.0, 0.85, 0.0}
\definecolor{cadmiumgreen}{rgb}{0.0, 0.5, 0.24}

\usepackage{amssymb,amsmath}
\usepackage{exscale}
\usepackage{makeidx,shortvrb,latexsym}
\usepackage{epstopdf}
\usepackage{tabularx, booktabs, multirow}
\usepackage[colorlinks=true,allcolors=NavyBlue]{hyperref}
\usepackage{color}
\journal{International Journal of Plasticity. Accepted for publication}

\begin{document}

\begin{frontmatter}

\title{Effect of slip transmission at grain boundaries in Al bicrystals}
\author[myaddress]{S.~Haouala}
\author[myaddress]{R.~Alizadeh}
\author[mythirdaddress]{T. R.~Bieler}
\author[myaddress,mysecondaryaddress]{J.~Segurado}
\author[myaddress,mysecondaryaddress,cor]{J.~LLorca }
\ead{javier.llorca@imdea.org}
\address[myaddress]{IMDEA Materials Institute, C/ Eric Kandel 2, 28906 - Getafe, Madrid, Spain}
\address[mythirdaddress]{Department of Chemical Engineering and Materials Science, Michigan State University, Michigan, USA.}
\address[mysecondaryaddress]{Department of Materials Science, Polytechnic University of Madrid, E. T. S. de Ingenieros de Caminos, 28040 - Madrid, Spain}

\cortext[cor]{Corresponding author.}

\begin{abstract}
The effect of slip transfer on the deformation mechanisms of Al bicrystals was explored using a rate-dependent dislocation-based crystal plasticity model. Three different types of grain boundaries (GBs) were included in the model by modifying the rate of dislocation accumulation near the GB in the Kocks-Mecking law, leading to fully-opaque (dislocation blocking), fully-transparent and partially-transparent GBs. In the latter, slip transmission is only allowed in pairs of SS in neighbour grains that are suitably oriented for slip transfer according to the Luster-Morris parameter. Modifications of the GB character led to important changes in the deformation mechanisms at the GB. In general, bicrystals with fully-opaque boundaries  showed an increase in the dislocation density near the GB, which was associated with an increase in the Von Mises stress. In contrast, the dislocation pile-ups and the stress concentration were less pronounced in the case of partially-transparent boundaries as the slip in one grain can progress into the next grain with some degree of continuity. No stress concentrations were found at these boundaries for fully-transparent boundaries, and there was continuity of strain across the boundary, which is not typical of most experimentally observed GBs \citep{HNP18, BAP19}. Simulations of ideal bicrystals oriented for favorable slip transfer on the most highly favored slip system in grains with high Schmid factors for slip transfer depends  on the number of active SS in operation in the neighborhood and that most boundaries will lead to nearly opaque conditions while some boundaries will be transparent. Finally, the model was applied to a particular experimentally observed GB in which slip transfer was clearly operating indicating that the model predicted a nearly transparent GB.

\end{abstract}

\begin{keyword}
 Grain boundaries \sep  Slip transfer \sep  Crystal plasticity 
\end{keyword}

\end{frontmatter}

\section{Introduction}

There has been extensive research for the past three decades to identify how the microstructure affects the deformation of polycrystalline materials. Length scale effects installed into homogenized continuum polycrystal material models can simulate the effects of grain size and evolving dislocation density \citep{AV06, VD09} but they cannot take into account the heterogeneous deformation that is commonly observed in polycrystals. An excellent example can be found in the insightful image of an initially polished copper multicrystal  deformed in tension by  \cite{DRC00}, which shows heterogeneous deformation of each grain via slip traces that differ significantly in each grain, as well as within different regions of the same grain. Such images challenge the standard continuum assumption of Taylor models where the strain of each grain is uniform, and slip arises from the five most favoured slip systems (SS) within each grain. 

Development of the viscoplastic self-consistent model \citep{LT93} enabled different grain orientations to be strained by different amounts due to the accommodation effects of softer grain orientations to the greater deformation resistance of harder grains (where slip can also occur on less than five SS in a given grain orientation). However, this approach does not consider the actual grain boundary (GB) neighbour interactions or the geometry of grains (refinements to overcome this limitation have been extensive \citep{L01, ZPL17}. Because each grain deforms according to the deformation processes taking place in neighboring grains as well as itself, each grain has complex evolving boundary conditions governing its deformation \citep{MD14}.

This complex set of boundary conditions makes approaches such as finite element modeling of mesoscale grain geometry necessary \citep{ZLE15}. To this end, crystal plasticity finite element modeling is effective because the deformation of each grain is constrained to occur by the available SS within each grain and such models have seen much attention in the past decade \citep{KSBD04, L01, SLL18}. Crystal plasticity finite element modeling of polycrystals (with realistic smooth and complex shaped boundaries) is commonly accomplished using two approaches for the evolution of the slip resistance \citep{SLL18}. in the phenomenological approximation, the shear rate of each slip system evolves in accordance with an analytical function of resolved shear  stress that also considers the activity of other SS in each integration point in an element using a hardening interaction matrix to model latent hardening \citep{REH10}. Alternatively, dislocation density-based approaches are used, such that the local strain history causes development of a dislocation density within each integration point \citep{LLK11, HSL18}.

While both of the above approaches generate heterogeneous deformation in different parts of a given grain that is similar to experimental observations, the agreement between such models and experimental measurements with the same grain sizes and textures has not been fully convincing.  For example, comparisons between experimental measurements of activated SS and simulations of the same (usually simplified) microstructure  often show good agreement about which SS are activated, but the agreement is worse in regions near GBs and triple points \citep{YWBEC11,LCBBW15}.  One reason for these differences is that the finite element model intrinsically provides no properties of GBs and heterogeneous deformation arises from the discontinuity in properties on either side of the boundary.  It is well known that GBs provide barriers for dislocations, but material models provide no way to harden the properties of a grain locally due to the presence of a GB unless such properties are built into the geometry of the microstructure and/or the constitutive model.  This requires introduction of length scales that are not intrinsically part of a finite element formulation.  Hence, the default condition of a finite element mesoscale simulation of a deforming microstructure assumes that GBs have no resistance to the transfer of strain. 

There is much experimental evidence of hardening effects at GBs. For example, early observations of bicrystal deformation show that SS other than the most favoured one are active in some cases, but only near a GB \citep{LC57}. In other cases, slip in one grain transfers directly onto a nearly aligned slip system in the neighboring grain \citep{SWC88,LRB90}. Experimental observations of polycrystals using transmission electron microscopy (TEM) also indicate that many boundaries act as barriers for slip, resulting in nucleation of limited amounts of slip on other systems within the same grain to accommodate the slip resistance effects of a GB.  When dislocations pile up at a GB, this results in an accumulation of dislocations of one sign, leading to localized lattice rotation near boundaries, which can also be detected with electron backscatter diffraction (EBSD) mapping and corresponding slip traces that terminate at GBs on both sides of the GB \citep{BEZ14,BZ17}.  However, slip traces are continuous across the boundary in other boundaries where slip transfer is observed, and there is little lattice rotation on either side of the boundary, while TEM images show continuous passage of dislocations through the boundary \citep{FA11, KECR14}.  It is assumed that the boundary  is essentially transparent to a dominant slip system and provides  little resistance to slip. Rules for these processes have been proposed \citep{LM95, LRB90} assessed with molecular dynamics \citep{SS14, W15} and, to some extent, with dislocation dynamics \citep{DC07}.  Hence, rather than using a statistical or homogenized approach to model the effects of GBs on mechanical properties with a length scale parameter, a deterministic approach based on physical observations should be possible to model the properties of GBs, and this is the motivation for the present investigation. 

Given that the finite element model intrinsically assumes transparent conditions at GBs, it is necessary to take into account the hardening effect of GBs. 
 Many models have been developed in recent years to account for dislocation density evolution with the ability to account for dislocation transport through the crystal. These models include continuum dislocation dynamics models \citep{HSZ14, H15, XE15} , mesoscopic field dislocation dynamics dynamics \citep{A01, BTD14}  or higher-order CP frameworks \cite{G08}. However, in all the cases, the underlying coupled partial differential equations governing the models are complex (transport problems coupled with elliptic partial differential equations) and their numerical treatment is complicated and numerically very expensive. Moreover, such models are in general devoted to intragranular deformation and the conditions for slip transmission between grains using such a higher order approach has been only considered in detail in \cite{G02}. The effect of grain boundaries can also be considered by means of strain gradient plasticity models, which take into account the geometrically necessary dislocations that arise from the deformation incompatibility between neighbour grains \citep{MARR06, AB00, EPB02, CBA05, BBG07, BER10}.  Moreover, \cite{AN07} developed a strain gradient plasticity  to account for grain boundaryÐgrain interior interactions. This approach was able to capture the effects of dislocation pileups and allowed the development of an analytical expression which predicts the critical stress at which dislocation transmission/emission takes place at a GB. Nevertheless, these approaches do not take explicitly into account the geometry of the SS and of the GB.
 
 Other approaches to account for GBs in the framework of CP include developing a mantle layer of elements in the GB, where the properties are either initially different from the grain interior, or evolve differently than in the grain interior, to provide the appropriate hardening effects \citep{PNB09}.  \cite{LLK11, LSF14} presented another approach in which the information from polycrystal simulations using a dislocation-based crystal plasticity model was used at another length scale to enforce local slip transmission criteria at the GBs depending on the orientation and on the GB strength. More recently, \cite{HSL18} and \cite{RHL19} have developed a dislocation-based model in which the rate of dislocation storage is not constant within the grain but increases as the distance to the GB decreases. This  results in dislocation accumulation at boundaries and hence, higher stresses develop near GBs.  Without any assumptions regarding slip behavior other than intrinsic slip velocity, the model was nearly able to match the flow stress of different FCC (Al, Ni Cu and Ag)  polycrystals as a function of the average grain size.  However, the hardening in the model was slightly higher than the experimental results for polycrystals with small average grain size ($<$ 20 $\mu$m) and it was assumed that  not accounting for slip transfer could account for these differences. Clearly, there are ways to model the discrete behavior of dislocations that are effective, but they are accomplished in computational settings that have idealized boundary conditions.  On the other hand, CP models can track evolving local stress states near grain boundaries, but this evolution is not likely to be accurate unless the local dislocation behavior is correctly modeled.  To make progress past this conundrum, the approach taken in this paper is to modify dislocation behavior near grain boundaries based upon simple physically motivated assumptions to determine if comparisons with physical experiments of the same microstructure lead to improvements.

In parallel, experimental investigations of slip transfer in Al polycrystals have provided a statistically relevant evidence that slip transfer across GBs took place when the Luster-Morris parameter $>$ 0.95, which corresponds to low-angle boundaries with $<$15$^\circ$ misorientation \citep{BAP19}. It is obvious that the presence or not of slip transfer through the GB has important consequences from the viewpoint of the SS activated in each grain and the build-up of stresses at the boundary (which may lead to the formation of ledges or to fracture). Moreover, constraint of the surrounding polycrystal or the presence of free surfaces may also influence the deformation mechanisms around the GB, but these phenomena have not been analyzed in detail. This is the main objective of this investigation, which is inspired by the experimental evidence of slip transfer across a GB in pure Al presented in section 2. The behavior of the GBs in various bycrystals - representative of the experimental conditions as well as of idealized conditions - is analyzed by means of dislocation-based crystal plasticity model in \citep{HSL18} and \citep{RHL19}, which was modified to take into account the effect of slip transfer along particular SS. 
It should be noted that this type of physically-based CP models based on classic continuum mechanics are  able to account for many of the phenomena of interest related to dislocation evolution in the grain such as slip, reproduction of hardening stages, thermal activation of deformation mechanisms, etc., in a simple manner.  In addition, computational homogenization models based on CP frameworks are now able to consider three dimensional polycrystalline models with millions of degrees of freedom.  For these reasons, in this paper we consider a classical framework with a simple physically-motivated crystal plasticity model.  The model can be used to analyze simple cases such as a bicrystal, and provide simulations of experiments, but in addition, this model can be further used in a full polycrystal without any other modifications. The modified model is presented in section 3, while the details of the numerical simulations are summarized in section 4. The simulation results of the behavior of the bicrystals under unconstrained and constrained deformation are presented and discussed in section 5, while the main conclusions of the paper are summarized in the last section.

\section{Experimental evidence of slip transfer across the GB}

\begin{figure}   \centering
 \includegraphics[scale=0.5]{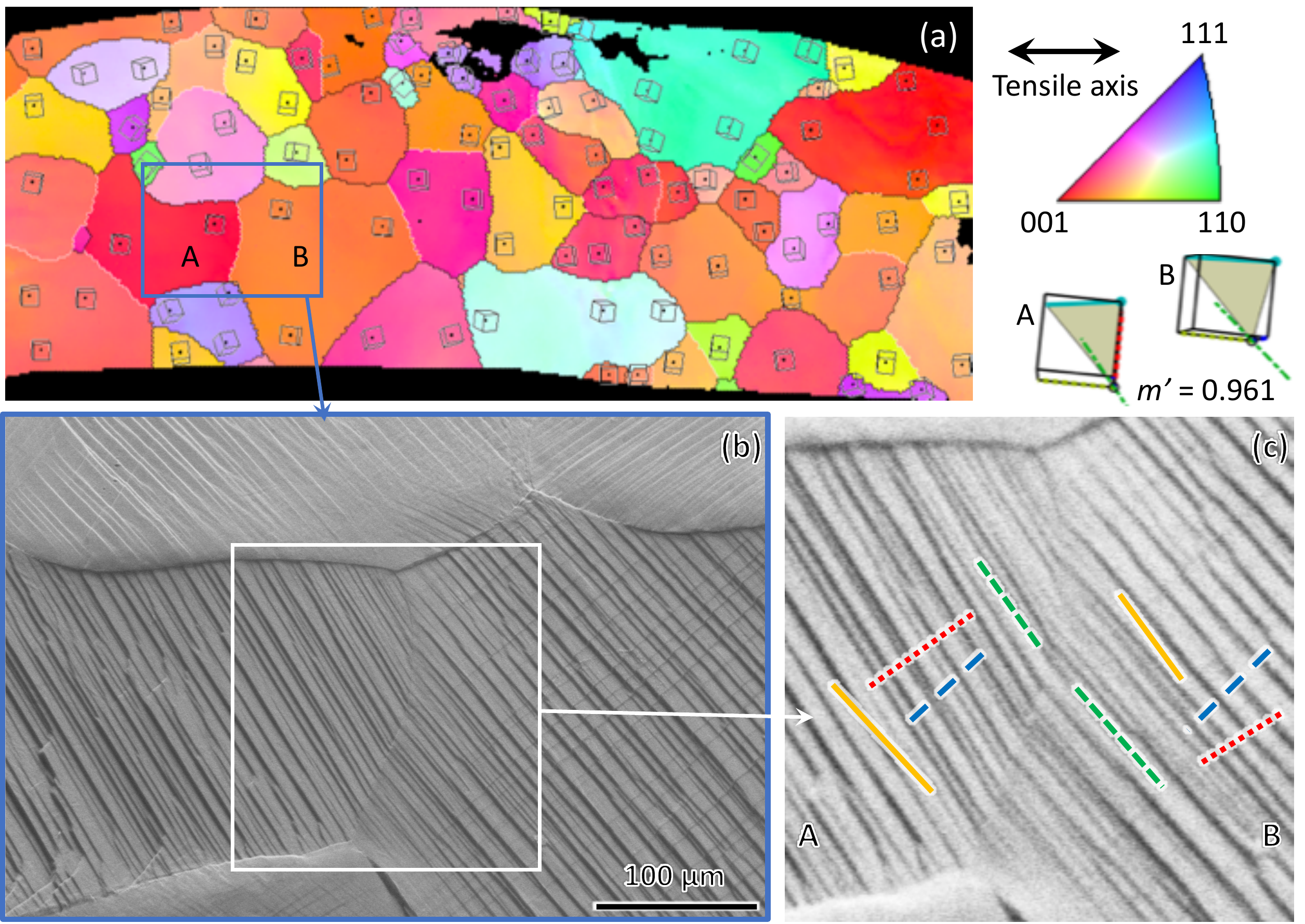}
\caption{(a) Crystal orientation map obtained by EBSD of the front surface of the Al foil tensile sample. The slip planes and slip directions corresponding to the slip system 1 in grains A and B (with $m'$ = 0.961) are shown. (b) Secondary electron micrograph of the deformed foil around the GB analysed in Table \ref{slipsystems}. (c) Detail of the slip traces around the GB. The green, red, blue and yellow lines correspond to the calculated traces of the planes for SS 1-3, 4-6, 7-9 and 10-12, respectively, on the surface of grain B to the left and grain A to the right of the GB.  The most apparently active SS are systems 1 on both sides, and 9 to a greater degree in the right grain (A) than the left grain (B; the slip trace topography for SS 9 is small, suggesting that the Burgers vector is nearly parallel to the surface).
}
\label{exp}
\end{figure}

Slip transfer across GBs was studied in annealed polycrystalline Al foils of 200 $\mu$m in thickness deformed in uniaxial tension up to 5\% strain \citep{BAP19}. The average grain size was 390 $\pm$ 30 $\mu$m  and the grain orientation was measured by EBSD prior to deformation. (Fig. \ref{exp}a). The GB analyzed is in the center of the outlined area marked in Fig. \ref{exp}a,b) and it is almost perpendicular to the horizontal deformation axis. After deformation, the sample was examined in the scanning electron microscope and the secondary electron micrographs of the region near the GB are shown in Fig. \ref{exp}b). The clearly visible slip traces identify the active SS operating in each grain. From the grain orientations (Table \ref{tab1}), the slip trace directions and Schmid factors (SF) in both grains across the GB were computed based upon the (convenient) assumption of uniaxial tension using the slip system definition in Table \ref{tab1-1}. Every slip system is listed in Table \ref{slipsystems} in descending order of the SFs corresponding to the 12 SS (SS) in grain B to the left of the boundary (first two columns) and the SFs corresponding to the 12 SS in grain A to the right of the boundary (first two rows). Because both grains have a near-cube orientation, there are eight SS with SFs above 0.3.  The underlined SS are those whose traces would not be highly visible on the specimen surface due to the Burgers vector being nearly parallel to the surface, so their activation cannot be confidently confirmed by slip trace analysis. The slip traces corresponding to the visible SS with the highest SFs are indicated in Fig. \ref{exp}c) with different colored line segments: green for SS 1-3, red for SS 4-6, blue for the SS 7 and yellow for the SS 9-12. The experimentally observed slip traces match very well with the green traces of SS1 in both grains. In addition, there are faint slip traces in both grains aligned with the blue trace. Because the GB is difficult to distinguish in Fig. \ref{exp}c), this indicates good compatibility in the deformation that does not lead to the formation of a ledge, indicating uniform strain in both grains near the boundary.

\begin{table}
\caption {Crystallographic orientation of the crystals across the GB for the ideal cases of single
and double slip as well as for the experimental GB in Fig. \ref{exp}. The Euler angles $(\phi_1, \Phi, \phi_2)$ are expressed in degrees.}\label{tab1}
\centering
   \begin{tabular}{ l c c c  }
   \hline
    Case  & Grain A & Grain B & Misorientation \\ 
    Single slip  & (11 25 13)  & (-11 -25 -13) & 57.7$^\circ$ \\ 
    Double slip  & (0 30 45) & (34 -30 -45 ) & 52.1$^\circ$\\ 
  Experimental  & (8.0 89.3 354) & (2.6 101.6 354) & 13.4$\circ$ \\ 
      \hline
   \end{tabular}
 \end{table}   

\begin{table}[]   \caption {Number, slip plane normal and slip direction of each one of the 12 slip planes of the FCC crystals }\label{tab1-1}
\centering
   \begin{tabular}{| c | c | c | c |}
   \hline
Number & Slip plane normal & Slip direction &\\
    \hline
    & & &\\
1 & $(\bar{1 } 1 1)$ &$[ 0 1\bar{1 } ]$ & 
\begin{tikzpicture}
\color{cadmiumgreen}\draw [thick] (1,0) -- (2,0);
\end{tikzpicture}\\
2 &  $ (\bar{1 } 1 1)$ & $[1 0 1]$ & \begin{tikzpicture}
\color{cadmiumgreen}\draw [thick,dashed] (1,0) -- (2,0);
\end{tikzpicture}\\
3 &  $ (\bar{1 } 11) $ & $[\bar{1 }\bar{1 } 0 ]$ &\begin{tikzpicture}
\color{cadmiumgreen}\draw [thick,dotted] (1,0) -- (2,0);
\end{tikzpicture}\\
4 & $(111)$ &$[ 0\bar{1 } 1 ]$& \begin{tikzpicture}
\color{red} [thick] \draw(1,0) -- (2,0);
\end{tikzpicture}\\
5 &  $ (111) $ & $[1  0\bar{1 }]$&\begin{tikzpicture}
\color{red}\draw [thick,dashed] (1,0) -- (2,0);
\end{tikzpicture}\\
6 & $ (111) $ &$[ \bar{1 } 1  0]$ & \begin{tikzpicture}
\color{red}\draw [thick,dotted] (1,0) -- (2,0);
\end{tikzpicture}\\
7 & $(\bar{1 }\bar{1 } 1)$ &$[ 0 1 1 ]$& \begin{tikzpicture}
\color{blue}\draw [thick] (1,0) -- (2,0);
\end{tikzpicture}\\
8 &  $ (\bar{1 }\bar{1 } 1) $ & $[\bar{1 } 0  \bar{1 }]$& \begin{tikzpicture}
\color{blue}\draw [thick,dashed] (1,0) -- (2,0);
\end{tikzpicture}\\
9 & $ (\bar{1 }\bar{1 } 1) $ & $[1 \bar{1 }0]$ &\begin{tikzpicture}
\color{blue}\draw [thick, dotted] (1,0) -- (2,0);
\end{tikzpicture}\\
10 &  $(1 \bar{1}1)$ & $[0\bar{1 }\bar{1 }]$& \begin{tikzpicture}
\color{chromeyellow}\draw [thick] (1,0) -- (2,0);
\end{tikzpicture}\\
11 &   $ (1 \bar{1 }1) $ & $[\bar{1}01]$&  \begin{tikzpicture}
\color{chromeyellow}\draw [thick,dashed] (1,0) -- (2,0);
\end{tikzpicture}\\
12 &  $ (1 \bar{1}1) $ &  $[110]$&  \begin{tikzpicture}
\color{chromeyellow}\draw [thick,dotted] (1,0) -- (2,0);
\end{tikzpicture}\\
     \hline
   \end{tabular}
 \end{table}  

\begin{sidewaystable}
\caption {$m'$ for different pairs of SS across the GB in \ref{exp}b), which is misoriented by 13.4$^\circ$. The first column and the first row indicate the number (N) of the SS in grains A and B, that are to the left and to the right of the GB in Fig. \ref{exp}b). The second row and the second column indicate the corresponding SF for each SS assuming uniaxial tension. The SS marked with bold font have a high $m'$ with the corresponding SS in the adjacent grain. The SS underlined would not provide highly visible traces on the specimen surface.}\label{slipsystems}
\centering
   \begin{tabular}{ l| l l || l  l  l l  l  l  l l l l l l }
      \hline
    \multicolumn{15}{c}{Slip system number (N)  and SF of grain B} \\
    \hline
 \multirow{15}{*}{\rotatebox{90}{\parbox{5cm}{\centering Slip system number (N) and  SF of grain A}}}& & \multicolumn{1}{ l|}{N}& $\mathbf{10}	$ & $ \mathbf{1}$ & $\mathbf{\underline{6}}	$ & $ \mathbf{\underline{3}}$ & $ \mathbf{\underline{12}}	$ & $\mathbf{\underline{9}}$ & $ \mathbf{4}$& \multicolumn{1}{ l|}{$ \mathbf{7}$} &$11$&$5$ &$2$ &$8$\\
& N & \multicolumn{1}{ l|}{SF }& $\mathbf{0.45}	$ & $ \mathbf{0.44}$ & $\mathbf{\underline{0.41}}	$ & $\mathbf{\underline{0.40}}$ & $ \mathbf{\underline{0.40}}$ & $\mathbf{\underline{0.39}}$ & $0.36$& 
\multicolumn{1}{ l|}{$0.36$}&$0.05$ &$0.04$ & $0.04$& $0.03$\\
   \cline{2-15}
&$\mathbf{\underline{9}}$&$\mathbf{\underline{0.47}}$ & $ 0.155$ & $0.073$ & $0.519$ & $0.004$ & $0.006$ & $\mathbf{\underline{0.970}}$ & $0.215$ & 
\multicolumn{1}{ l||}{$0.547$}& $0.161$ & $0.304$ & $0.077$ & $0.423$\\
&$\mathbf{\underline{12}}$& $\mathbf{\underline{0.46}}$ & $ 0.627$ & $ 0.150$ & $	0.000$ & $0.429$ & $\mathbf{\underline{0.967}}$ & $0.000$ & $0.066$ & 
\multicolumn{1}{ l||}{$0.227$} & $0.340$ & $0.066$ & $0.279$ & $0.227$\\
&$\mathbf{1}$&$\mathbf{0.42}$ & $0.053$ & $ \mathbf{0.961}$ & $0.151$ & $0.615$ & $0.149$ & $0.278$ & $0.256$ & \multicolumn{1}{ l||}{$0.108$} & $0.096$ & $0.106$ & $0.346$ & $0.170$ \\
&$\mathbf{10}$&$0.40$ & $\mathbf{0.962}$ & $ 0.096$ & $0.081$ & $0.146$ & $0.330$ & $0.150$ & $0.042$ & 
\multicolumn{1}{ l||}{$0.348$ }& $0.633$ & $0.123$ & $0.243$ & $0.198$ \\
&$\mathbf{7}$&$0.34$ & $ 0.270$ & $ 0.039$ & $	0.220$ & $0.059$ & $0.092$ & $0.411$ & $0.115$ & \multicolumn{1}{ l||}{$\mathbf{0.950}$}& $0.177$ & $0.334$ & $0.098$ & $0.540$  \\
&$\mathbf{4}$&$0.32$ & $ 0.105$ & $ 0.361$ & $0.559$ & $0.231$ & $0.295$ & $0.077$ & $\mathbf{0.952}$ & \multicolumn{1}{ l||}{$0.030$} & $0.190$ & $0.393$ & $0.130$ & $0.047$ \\
&$\mathbf{\underline{3}}$&$\underline{0.31}$ & $0.151$& $ 0.338$& $0.000$& $\mathbf{\underline{0.965}}$& $0.233$& $0.00$& $0.090$& \multicolumn{1}{ l||}{$0.308$}& $0.082$& $0.090$& $0.627$& $0.308$\\
&$\mathbf{\underline{6}}$&$0.30$ & $0.265$&$0.153$&$\mathbf{\underline{0.971}}$&$0.008$&$0.010$&$0.133$&$0.403$&\multicolumn{1}{ l||}{$0.075$}&$0.275$&$0.075$&$0.075$&$0.075$\\
 \cline{2-11}
&$8$&\multicolumn{1}{ l|}{$0.13$} &$ 0.114$ &$ 0.112$ & $ 0.299$& $0.063 $& $0.098 $& $ 0.560$&$ 0.330$ & $ 0.403$ & $0.016 $ & $ 0.0136$ & $ 0.175$ & $ 0.963$ \\
&$2$&\multicolumn{1}{ l|}{$0.11$} &$ 0.098$ & $ 0.623$& $0.151 $& $0.350 $& $0.085 $&$ 0.278$ &$ 0.166$ &$ 0.200$ &$0.014 $ &$ 0.015$ &$0.973 $ &$ 0.478$ \\
&$11$&\multicolumn{1}{ l|}{$0.05$} &$ 0.335$ & $ 0.247$& $0.081$& $0.283 $&$0.637 $&$ 0.151$ &$ 0.108$ & $ 0.121$  & $0.972 $  & $ 0.189$  & $ 0.036$  & $ 0.030$ \\
&$5$&\multicolumn{1}{ l|}{$0.03$} &$0.160 $&$0.208 $&$ 0.412$&$0.239 $&$0.305$ &$ 0.057$&$ 0.549$& $0.045$& $0.465 $& $0.961 $& $ 0.031$& $ 0.011$\\
  \hline
   \end{tabular}
 \end{sidewaystable}  

\begin{figure}  
 \centering
 \includegraphics[scale=0.7]{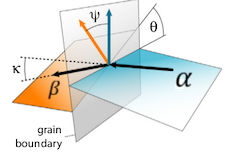}
\caption{Angles and vectors used to evaluate the likelihood of slip transfer across a GB from slip system $\alpha$ to $\beta$ \citep{BERKMCR09}.\label{GeoST}}
\end{figure}

The geometry of slip transfer between two SS $\alpha$ and $\beta$ on either side of a boundary is defined in Fig. \ref{GeoST} by the three angles: $\kappa$  (the angle between slip vectors), $\psi$ (the angle between slip plane normals), and $\Theta$ (the angle between the two slip plane intersections with the GB plane). Different criteria for slip transfer have been proposed in the literature \citep{LRB90, BMR16}, assuming that transmission is more likely when the slip plane and slip direction are closely aligned, and the angle between the slip planes in the boundary is small. Other important factors are minimizing the residual Burgers vector content left within the boundary after transmission, and a sufficiently high resolved stress to drive the transmitted slip. Recent experimental observations in Ti  \citep{HNP18} and Al  \citep{BAP19} have found a good correlation between slip transmission and the Luster-Morris parameter $m'_{\alpha\beta}$, which is expressed as

\begin{equation}
m'_{\alpha\beta} = \cos \psi \cos \kappa.
\end{equation}\label{LM}

\noindent An additional advantage of the $m'_{\alpha\beta}$ parameter is that it can be determined for each GB from the Euler angles of the adjacent grains \citep{BAP19}. In particular, \cite{BAP19} found that slip transfer in cube-oriented polycrystals was consistently found when $m'_{\alpha\beta} >$ 0.95, which corresponds with misorientations less than $\approx$ 18$^\circ$.

In the case of the GB in Fig. \ref{exp}, the misorientation angle is 13.4$^\circ$, and a consequence of this is that every slip system in grain A has a $m'_{\alpha\beta}$  parameter $>$ 0.95 with its corresponding slip system in grain B.  Thus, slip transfer across this GB is favored on any slip system, but there is convincing slip transfer through the grain boundary on one slip plane, as evident by the continuous slip traces and the lack of ledges along the GB (which are difficult to identify in Fig. \ref{exp}c). With higher misorientations, slip transfer is usually restricted to one pair of SS, which cannot fully accommodate the anisotropic deformation of each grain \citep{HNP18, BAP19}. The influence of slip transfer on the activation of the SS in each grain and on the build-up of stresses at the GB is not known, but there is some evidence for a threshold for slip transfer based upon the sum of the SFs times $m'_{\alpha\beta}$  for the two correlated SS  \citep{BAP19}.

\section{Crystal plasticity model including slip transfer at GBs}\label{sec1}

The mechanical behavior in each crystal within the polycrystal follows the dislocation-based crystal plasticity model presented in \cite{HSL18}, which is described sufficiently to motivate the modifications made to track slip transfer. The relationship between the resolved shear stress $\tau^{\alpha}$ acting on the slip system $\alpha$ and the corresponding plastic strain rate, $\dot{\gamma}^{\alpha}$, is expressed as \citep{K75, KL78})

\begin{equation}
\dot{\gamma}^{\alpha}=\dot{\gamma}_0\left(\frac{|\tau^{\alpha}|}{\tau^{\alpha}_c}\right)^{\frac{1}{m}} \operatorname{sgn}(\tau^{\alpha}),
\end{equation}

\noindent where $m$ is the strain-rate sensitivity coefficient, $\dot{\gamma}_{0}$ the reference shear strain rate and $\tau_{c}^{\alpha}$ the critical resolved shear stress (CRSS) on the slip system $\alpha$. Following \cite{T34}, the CRSS depends on the dislocation density on the different SS according to \citep{FBZ80}

\begin{equation}\label{tauc}
\tau^{\alpha}_c= \mu b\sqrt{\sum_{\beta}a^{\alpha \beta} \rho^{\beta}},
\end{equation}

\noindent where $\mu$ and $b$ denote the shear modulus and the Burgers vector, respectively, and $\rho^\beta$ is the dislocation density in the slip system $\beta$. The dimensionless coefficients $a^{\alpha\beta}$ of the dislocation interaction matrix represent the average strength of the interactions between dislocations in pairs of SS and can be determined by means of dislocation dynamics simulations for each type of crystal lattice \citep{DHK08, BCB13}.

The overall hardening of the crystal during deformation is controlled by the evolution of the dislocation density, which depends on the balance between the generation and the annihilation of dislocations \citep{KM03, T97}. This model implicitly assumes that there are enough dislocation sources in any location that they can be activated with a sufficiently large resolved shear stress.  The accumulation rate of dislocations in each slip system $\alpha$,  $\dot{\rho}^{\alpha}$, can be expressed as \citep{HSL18}

\begin{equation}\label{evolution-dislocation}
\dot{\rho}^{\alpha}=\frac{1}{b} \left(\max \left(\frac{1}{\ell^{\alpha}},\frac{K_{s}}{d_b}\right)- 2 y_c \rho^{\alpha} \right)| \dot{\gamma}^{\alpha}| ,
\end{equation} 

\noindent where $\ell^{\alpha}$ is the dislocation mean free path along the system $\alpha$, $d_b$ the distance to the closest GB along the slip system $\alpha$ and $y_c$ the effective annihilation distance between dislocations. Dislocations are generated as a result of the interaction of dislocations with other dislocations within the bulk of each crystal and at the GBs, that block dislocation slip and lead to the formation of pile-ups. The former mechanism depends on $\ell^{\alpha}$, which can be obtained as \citep{AS78, KDH08}

\begin{equation}\label{MFPd}
\ell^{\alpha}=\frac{K}{\sqrt{\displaystyle\sum_{\beta\neq\alpha}{\rho^{\beta}}}},
\end{equation}

\noindent where $K$ is the similitude coefficient relates the flow stress  with the average wavelength of the characteristic dislocation pattern and was estimated by \cite{SK11} for different FCC metals. The accumulation of dislocations at the GBs assumes that no dislocation can cross the boundary between crystals leading to a dislocation pile-up and to a local stress concentration, without any assumptions regarding slip behavior. This accumulation depends on the constant $K_s \approx 5$ that controls the storage of dislocations on the GB and was determined by means of dislocation dynamics simulations  \citep{SDK10}. The annihilation of dislocations in eq. \eqref{evolution-dislocation} is determined by $y_c$, the effective annihilation distance between dislocations, which can be taken as the average between the annihilation distance between edge dislocations (of the order of 6$b$) and screw dislocations (which depends on the stacking fault energy, temperature and strain rate) \citep{RHL19}.

Eq. \eqref{evolution-dislocation} assumes that slip transfer is always blocked at the GBs, which contradicts the experimental evidence (Fig. \ref{exp}). To include the possibility of slip transmission, this equation  was modified to account for slip transfer along pairs of SS with high $m'_{\alpha\beta}$. Thus, the $m'_{\alpha\beta}$ values for the slip system $\alpha$ with respect to all the SS $\beta$ in the closest neighbour grain were calculated from the orientation of both grains. If $m'_{\alpha\beta}$ was higher than a critical value (taken as 0.95 in this case), it was assumed that slip transfer was possible along this boundary and the term $K_{s}/{d_b}$ was not included in eq. \eqref{evolution-dislocation} for slip system $\alpha$ in this grain. Correspondingly, the same modification was introduced in the slip system $\beta$ of the grain across the boundary. It should be noted that these changes in the slip behavior are only active near the grain boundary according to eq. \eqref{evolution-dislocation} and do not modify the behavior of the bulk crystals.    Furthermore, there are no other grain boundary details present in the model such as structural or energy evolution with changing crystal orientation or evolving local stress states.

\section{Numerical model}\label{sec2}

\begin{figure}  
 \centering
\includegraphics[scale=0.6]{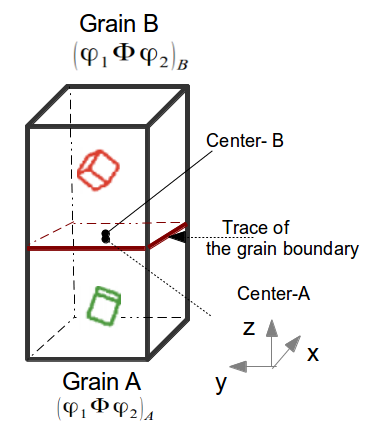}
 \includegraphics[scale=0.7, angle= 0]{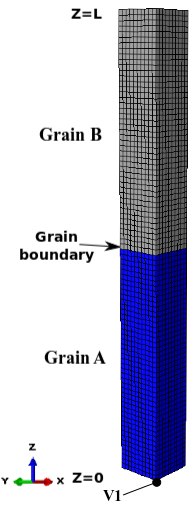}
\caption{(a) Schematic of the bicrystal showing points A and B adjacent to the GB, where strain history is plotted in Figs.  \ref{SS-shear-FS}, \ref{DS-shear-FS}, \ref{SS-shear-PBC},  \ref{DS-shear-PBC}, and \ref{Shear-exp}. (b)  Discretized finite element model of the bicrystal. The dimensions of each half of the bicrystal are $ 0.04 \times 0.04 \times 0.18$  mm$^3$  and each is discretized with $10 \times 10 \times 45 $ elements.\label{RVE}}
\end{figure}

The mechanical behaviour of several idealized Al bicrystals with a GB perpendicular to the loading axis and a total length $L=0.36$ mm, was analyzed using the finite element method. The schematic of the bycrystal is shown in Fig. \ref{RVE}a), while the discretized geometrical model  is presented in Fig. \ref{RVE}b) and includes $10 \times 10 \times 90 $ $3D$ solid elements or voxels (C3D8 elements in Abaqus/Standard with 8 nodes at the voxel corners and full integration). Thus, the length of each voxel was 8 $\mu$m.

The mechanical behavior of each grain was dictated by the rate-dependent crystal plasticity model in the context of finite strain plasticity (more details in \cite{HSL18}) which was modified to account for the possibility of slip transfer across the GB. The model parameters for Al deformed under quasi-static loading conditions were identified in \cite{RHL19} and are shown in Table \ref{tab:Al}. They include the elastic constants of the Al single crystals \citep{L06, KH15}, the viscoplastic parameters $\dot{\gamma}_0$ and $m$ \citep{EBG04}, and all the parameters that determine the interactions among dislocations and dislocations and GBs \citep{RHL19}.
It should be noted that the grain boundaries are not modeled as a single row of elements in each crystal in the bicrystal model. All the Gauss points in the finite element model for which the condition $\frac{1}{\ell^{\alpha}}  > \frac{K_{s}}{d_b}$  is fulfilled are affected by the presence of the grain  boundary, regardless of whether they are in the first row of elements in contact with the grain boundary or farther away. Moreover, as $\ell^{\alpha}$ varies during deformation in each Gauss point, the region affected by the grain boundary also changes during the simulation.

\begin{table}[ tp ] 
\caption {Parameters of the dislocation-based crystal plasticity model for Al single crystals}\label{tab0}
\begin{center}
   \begin{tabular}{ l  l  l  l l}
   \hline
 {\it Elastic constants} (GPa) & $C_{11}$= 108  & $ C_{12}$ = 61.3   & $C_{44}$= 28.5 & \\ 
     Shear modulus  (GPa) & $\mu$ = 25 & & & \\
     \hline
      {\it Viscoplastic parameters} & & \\ 
     reference shear strain rate   (s$^{-1}$) & $\dot{\gamma}_0 =0.001$  & & \\ 
     Strain rate sensitivity coefficient & $m$ = 0.05 & &\\
     \hline
     {\it Dislocation parameters}:\\
      Burgers vector (nm)  &$b$ = 0.286 & & \\
     Annihilation distance (nm)  & $y_c$ = 56 &  &\\
      \hline
      {\it Interaction coefficients:} & & & \\
      Self interaction &$0.122 $ &\\
      Coplanar interaction &$0.122 $ &\\
      Collinear interaction  &$0.657$&\\      
      Glissile junction &$0.137$& \\
      Hirth lock & $0.084$&\\
      Lomer-Cottrell  lock & $0.118$	&\\
    \hline
     Similitude coefficient  & $K$ = 9 & \\
     GB storage  & $K_s$ = 5 & \\
     \hline
      \hline
   \end{tabular}
 \end{center}\label{tab:Al}
 \end{table}   
 
Three different pairs of bicrystals were analyzed. The first two cases are idealized simulations to understand the effect of slip transfer with two contrasting cases where the mechanisms of deformation around the boundary involved a single slip system and two SS. In each case, two different sets of boundary conditions were considered. The first one was unconstrained deformation, in which the bicrystal was deformed in tension along the $z$ axis while the lateral surfaces were stress-free. The second boundary condition corresponded to constrained deformation: the bicrystal was deformed in tension along the $z$ axis while periodic boundary conditions were applied on the surfaces perpendicular to the $x$ and $y$ axes. In all cases, the nodes at $z$ = $L$ were displaced at a constant velocity along the $z$ axis, leading to a  constant strain rate of 7.0 10$^{-4}$ s$^{-1}$, up to a far-field applied strain of 2\%. The third case analyzed corresponds to the two crystals with the orientation in Fig. \ref{exp} to show that the crystal plasticity model with slip transfer is able to reproduce the experimental results. 

For each bicrystal and boundary conditions set, three simulations were carried out with different properties for the GB. In the first case (fully transparent GB), the term $K_{s}/{d_b}$ in eq. \eqref{evolution-dislocation} was not included in the constitutive equation of each grain and, thus,  slip in one grain can progress freely into the next grain through the boundary in transparent GBs. This is the default behavior considered in standard crystal plasticity simulations. The second case (fully opaque GB) includes the term $K_{s}/{d_b}$ in eq. \eqref{evolution-dislocation} in all the Gauss points of the numerical model, following \cite{HSL18} and \cite{RHL19}. Thus, dislocation pile-ups are formed on all SS at the GB and, thus, opaque GBs block dislocation motion and do not allow slip transfer through the boundary.  The partially transparent GB is found between these two bounding cases. The magnitude of the Luster-Morris parameter for each pair of SS in both crystals, $m'_{\alpha\beta}$, was determined; if $m'_{\alpha\beta}$ $>$ 0.95, the term $K_{s}/{d_b}$ is not included in the eq. \eqref{evolution-dislocation} corresponding to the SS $\alpha$ and $\beta$ in each grain. Thus, slip transfer is enabled for these pairs of SS but not for other pairs.

The finite element simulations of the bicrystal behavior were carried out in Abaqus/Standard \citep{A17} within the framework of the finite deformations theory with the initial unstressed state as reference.  The non-linear constitutive equations were integrated using a Newton-Raphson algorithm \citep{HSL18}. The constitutive equations developed in the previous section include the distance to the nearest GB for each slip system.  This information was computed and stored at the beginning of the simulations for each slip system in every Gauss point. The deformation gradient in these simulations was small and the distance from the Gauss point to the nearest GB did not change significantly during the analysis. Moreover, it was checked that the voxel discretization of the bicrystal was fine enough to obtain results independent of the voxel dimensions.

\section{Results and discussion}\label{sec3}

The behavior of three bicrystals with different orientations were analyzed using the crystal plasticity model presented above. The first two cases are ideal bicrystals.  In the first case, referred to hereafter as the single slip case, each crystal is oriented to have the maximum possible Schmid factor of 0.5, with these slip systems aligned to be perfectly aligned with each other.  In the second case, both crystals are oriented such that two slip systems have the highest possible Schmid factor of 0.43, but only one of them is perfectly aligned with one of the two favored slip systems in the other grain, referred to hereafter as the double slip case. Simulations of the ideal bicrystals were carried out using unconstrained and periodic boundary conditions, and they are analyzed separately. The third case corresponds to the experimental GB in Fig. \ref{exp} assuming unconstrained boundary conditions.

The crystallographic orientation of each crystal in the three bicrystals are given in Table \ref{tab1} using the Bunge Euler angle notation $(\phi_1, \Phi, \phi_2)$. The number, slip plane normal and slip direction of each one of the 12 slip sysytems of the FCC crystals is indicated in Table  \ref{tab1-1}.

\subsection{Idealized GBs} 

The magnitude of the SFs under uniaxial tension and of the Luster-Morris $m'_{\alpha\beta}$ parameter for each pair of SS is presented in Table \ref{tab2} for the bicrystal oriented for single slip. Similar information is presented in Table \ref{tab3} for the bicrystal oriented for double slip. In both tables, the seven highest SFs among all potential SS in grain B and grain A are assembled in descending order in the left column and the upper row, respectively. The numbers in the table indicate the values of $m'_{\alpha\beta}$ for each pair of SS. The $m'_{\alpha\beta} >$ 0.95 are printed in bold font. There is only one pair of SS with high SFs and high $m'_{\alpha\beta}$ in Table \ref{tab2}, indicating that this bicrystal is suited for single slip. In the case of Table \ref{tab3}, there are two SS in each grain with high SFs (2 and 10 in grain A and 8 and 4 in grain B), so both crystals are oriented for double slip. However, slip transfer (according to the Luster-Morris parameters in the table) is only possible between SS 2 and 8 but not between 10 and 4. There are also high $m'_{\alpha\beta}$ values for SS with low SF which are not expected to be active. 

\begin{sidewaystable}
\caption {$m'$ for different pairs of SS across the GB of two antisymmetrically oriented crystals (misorientation of 57.7$^\circ$) where one slip system is favored over others with the maximum SF and they are aligned nearly perfectly with each other. The first column and the first row indicate the number (N) of the slip system in the grains B and A, that are to the left and to the right of the GB in Fig. \ref{RVE}a). The second row and the second column indicate the corresponding Schmid factor (SF) for each slip system assuming uniaxial tension. The SS marked with bold font have a high $m'$ with other slip system from the adjacent grain.}\label{tab2}
\centering
   \begin{tabular}{l | l l | l  l  l l  l  l  l }
   \hline
    \multicolumn{10}{c}{Slip system number (N) and SF of grain A} \\
   \hline
 \multirow{9}{*}{\rotatebox{90}{\parbox{3,5cm}{\centering Slip system number (N) and  SF of grain \textbf{B}}}}& & N & $\mathbf{ 2}$ & $5$ & $ 10$ & $ 4$ & $ \mathbf{3}$ & $ \mathbf{1}$ & $ 7$\\ 
& N & SF & $\mathbf{0.5}$ & $0.46$ & $0.31$ & $0.28$ & $0.25$ & $0.24$ & $ 0.21$\\ 
  \cline{2-10}
& $\mathbf{8}$&$\mathbf{0.5} $  & $\mathbf{0.99} $ & $ 0.00$ & $ 0.16$ & $0.14$ & $ 0.53$ & $0.46$ & $ 0.18$  \\ 
  & $11$&$0.46$  & $0.00$ & $ 0.20$ & $0.65 $ & $ 0.28$ & $0.15 $ & $0.15 $ & $ 0.43$ \\ 
  &$ 4$&$0.31$  & $0.16$ & $0.65 $ & $0.42$ & $0.39$ & $ 0.00$ & $0.16$ & $ 0.49$ \\
 & $10$&$0.28 $  & $ 0.14$ & $0.28$ & $ 0.39$ & $0.30$ & $ 0.32$ & $0.17$ & $0.26$ \\ 
 & $ \mathbf{9}$&$0.25$  & $0.53$ & $0.15$ & $0.00$ & $0.32$ & $ 0.46$ & $\mathbf{0.99} $ & $ 0.00$\\ 
  &$\mathbf{7}$&$0.24 $  & $ 0.46$ & $ 0.15$ & $ 0.16$ & $ 0.17$ & $\mathbf{ 0.99}$ & $ 0.53$ & $0.18 $ \\ 
  &$ 1$&$0.21$  & $ 0.18$ & $0.43$ & $ 0.49$ & $ 0.26$ & $0.00$ & $ 0.18$ & $ 0.63$ \\ 
      \hline
   \end{tabular}
\end{sidewaystable}
 
\begin{sidewaystable}
\caption{$m'$ for different pairs of SS across the GB of two  crystals where two SS have the same maximum SF, but only one set is aligned with each other (misorientation of 52.1$^\circ$).
The first row and the first column indicate the number (N) of the slip system in the grains A and B, that are to the left and to the right of the GB in Fig. \ref{RVE}a). The second row and the second column indicate the corresponding Schmid factor (SF) for each slip system assuming uniaxial tension. The SS marked with bold font have a high $m'$ with other slip system from the adjacent grain.}\label{tab3}
\centering
   \begin{tabular}{ l| l l | l  l  l l  l  l  l }
      \hline
    \multicolumn{10}{c}{Slip system number (N) and SF of grain A} \\
   \hline
 \multirow{9}{*}{\rotatebox{90}{\parbox{3,5cm}{\centering Slip system number (N) and  SF of grain B}}}& & N& $\mathbf{	2}	$ & $ 10$ & $\ 5	$ & $ 4$ & $ 	3	$ & $12$ & $ 	\mathbf{1}$\\
& N & SF & $\mathbf{0.43}	$ & $ 0.43	$ & $0.32	$ & $0.32$ & $ 0.25	$ & $0.25$ & $0.18$\\
   \cline{2-10}
&$4$&$0.43$ & $ 0.19$ & $ 0.42$ & $0.57$ & $0.29$ & $ 0.04$ & $ 0.06$ & $0.14$\\
&$\mathbf{8}$&$0.43$ & $\mathbf{0.98}$ & $ 0.19$ & $ 0.03$ & $ 0.13$ & $ 0.61$ & $ 0.22$ & $ 0.37$\\
&$11$&$0.32$ & $0.03$ & $	0.57$ & $0.18$ & $0.40$ & $	0.18$ & $0.37$ & $0.21$\\
&$10$&$0.32$ & $	0.13$ & $0.29$ & $0.40$ & $0.41$ & $0.35$ & $0.73$ & $0.22$\\
&$6$&$0.25$ & $0.22$ & $0.06$ & $0.37$ & $0.73$ & $0.13$ & $0.17$ & $0.36$\\
&$\mathbf{9}$&$0.25$ & $0.61$ & $0.04$ & $0.18$ & $0.35$ & $0.37$ & $0.13$ & $\mathbf{0.98}$\\
&$5$&$0.18$ & $0.03$ & $0.36$ & $0.19$ & $0.43$ & $0.18$ & $	0.23$ & $0.21$\\
  \hline
   \end{tabular}

\end{sidewaystable}

\subsubsection{Unconstrained boundary conditions}
\begin{figure} 
\centering
\begin{subfigure}{.47\textwidth}
  \centering
  \includegraphics[width=\linewidth]{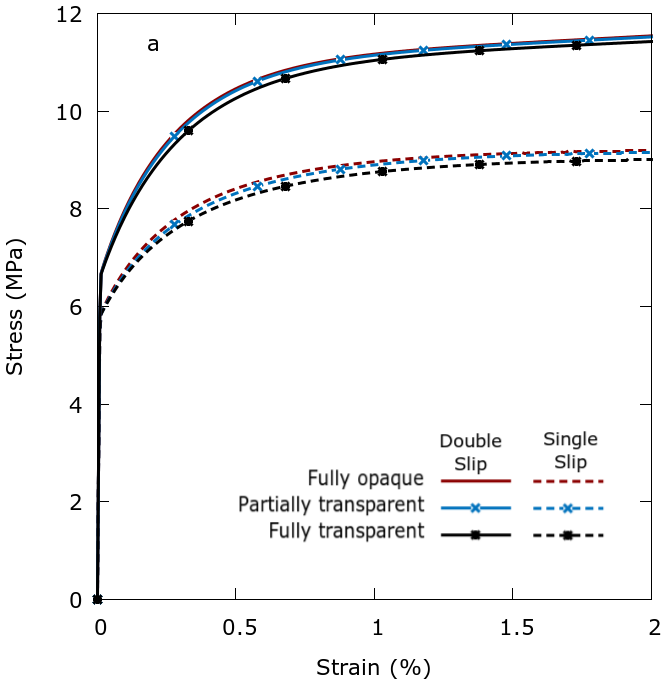}
  \caption{ }
  \label{fig:sub1}
\end{subfigure}%
\begin{subfigure}{.5\textwidth}
  \centering
  \includegraphics[width=\linewidth]{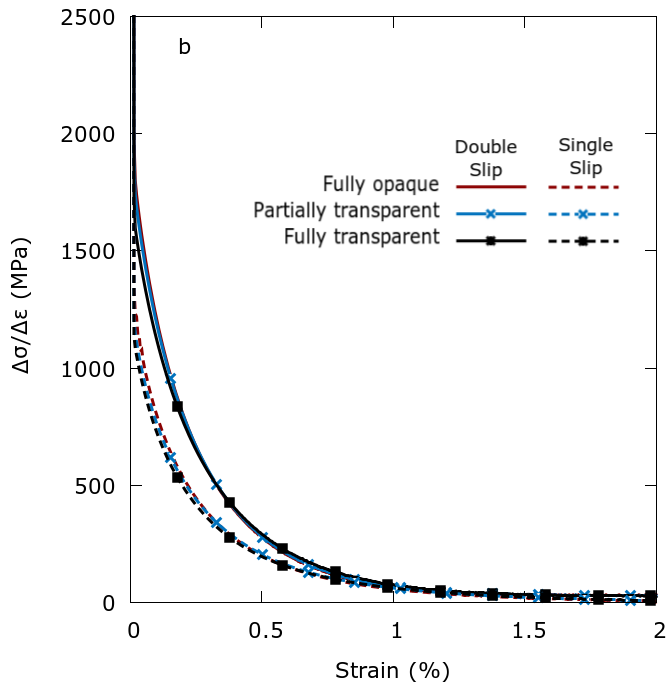}
  \caption{ }
  \label{fig:sub2}
\end{subfigure}
\caption{(a) Stress-strain curves and (b) strain hardening rate-strain curves under unconstrained boundary conditions for the single-slip (dashed line) and double-slip (solid line) Al bicrystals as a function of the GB characteristics.  
}
\label{FS-StressStrain}
\end{figure}

The tensile stress-strain behaviour near the GB of the two Al bicrystals computed with unconstrained boundary conditions, and the corresponding hardening rate-strain curves are plotted in Figs. \ref{FS-StressStrain}a) and b), respectively. For both bicrystals, the results of the simulations show a higher flow stress with decreasing GB transparency although the differences were small.  The strengthening induced by GBs results from the region near the GB in which the storage of dislocations at the boundary increased the dislocation density. The hardening rate of the single-slip bicrystals (Fig. \ref{FS-StressStrain}a) is always lower than those of the double-slip bicrystals (Fig. \ref{FS-StressStrain}b) because of latent hardening as the  two most stressed SS in the latter are found in intersecting planes. The evolution of shear strain at the center of the sample near the GB in the lower grain A and the upper grain B (as indicated in Fig. \ref{RVE}a) is shown in Fig. \ref{SS-shear-FS} and Fig. \ref{DS-shear-FS} for the single- and double-slip Al bicrystals, respectively, for fully transparent, partially transparent, and fully opaque boundaries.

\subsubsection{Single-slip bicrystal}

\begin{figure} 
 \centering
 \includegraphics[width=\linewidth]{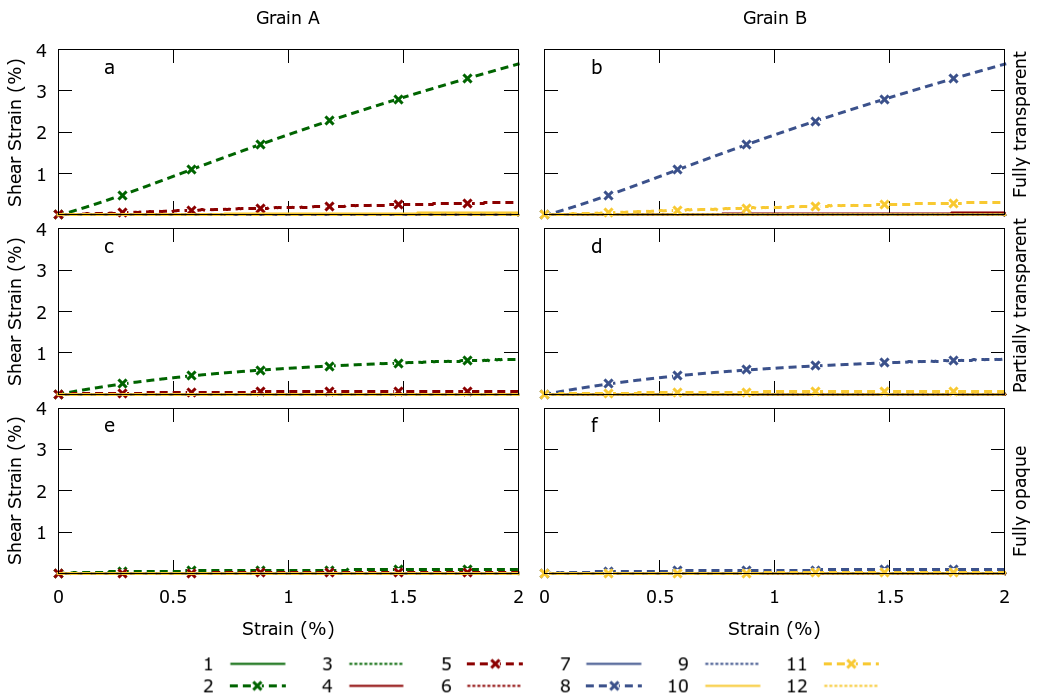}
\caption{Accumulated shear strain in each slip system at the center of the GBs for a single-slip-type Al bicrystal as a function of the far-field applied strain  under unconstrained boundary conditions.  The amount of shear on the favored SS decreases with increasing opacity.
}\label{SS-shear-FS}
\end{figure}

For the single-slip favored Al bicrystal, the deformation of the two crystals in the presence of a transparent boundary is dominated by the slip system with the highest SF in each crystal as noted in bold in Table \ref{tab2} and Figs. \ref{SS-shear-FS}a) and b) (system 2 in crystal A and system 8 in crystal B). In Table \ref{tab2}, the rows and columns are sorted in the order of decreasing SF, such that the SS that are most likely to be active are in the upper left corner of the table.  The body of the table has the $m'_{\alpha\beta}$  values that correspond with each pair of SS along the top and left edge.  Figs. \ref{SS-shear-FS}a) and b) show dominant accumulated shear on the two favored SS (system 2 in crystal A and system 8 in crystal B). Moreover,  a secondary SS (5 in crystal A and 11 in crystal B also noted in bold) starts to produce plastic strain when the far-field applied strain was higher than 0.25\%. 

The accumulated shear strains in each slip system near the boundary in Figs. \ref{SS-shear-FS}c to f) show that the strains of all SS decrease as the GB changes from fully transparent to partially transparent to fully opaque. In the partially transparent case, the near-perfect alignment between slip system 2 in grain A and slip system 8 in grain B ($m'=0.99$, Tab. \ref{tab2}) enables slip transmission on this slip system in both crystals, but the GB effect of limiting slip is shown on the other less-active SS (Fig. \ref{SS-shear-FS}c and d). So, by limiting slip transfer to only the most preferred SS in each grain, the amount of shear strain decreased in all active SS near the boundary.  On the most favored slip system, the shear strain decreased significantly to about 22\% of the transparent condition (from about $0.037$ to $0.022$). Nevertheless, the shear strain decreased much more when the GB was opaque (Fig. \ref{SS-shear-FS}e and f). This same overall effect was observed consistently in many other locations near the boundary and on the free surface as well, but with some variability from location to location.

\subsubsection{Double-slip bicrystal}

\begin{figure} 
 \centering
 \includegraphics[width=\linewidth]{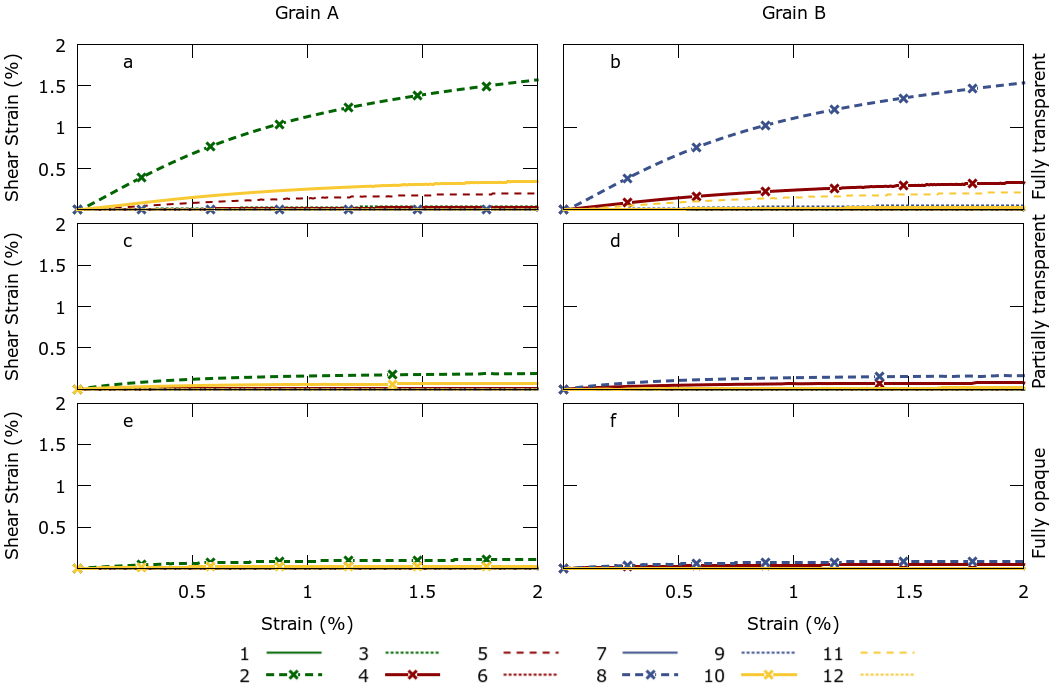}
\caption{Accumulated shear strain in each slip system at the center of the GBs for a double-slip-type Al bicrystalas a function of the far-field applied strain under unconstrained boundary conditions.  The decrease in shear strain due to partial transparence is greater than the single slip bicrystal in Fig. \ref{SS-shear-FS}. 
}\label{DS-shear-FS}
\end{figure}

For the double-slip bicrystal, the accumulated shear strain in each slip system is plotted in Fig. \ref{DS-shear-FS} as  a function of the far-field applied strain for different types of boundaries.  
In the case of the transparent boundary (Fig. \ref{DS-shear-FS}a) and b), slip transfer is mainly concentrated  between slip system 8 in grain B with slip system 2 in grain A ($m'=0.98$), and there is significantly more shear on these two SS than any other.  SS 10 in grain A and 4  in grain B have the same high SF (0.43), but their activation is much lower. Moreover, SS 5 in grain A and 11 in grain B, which also have the same SF (0.32), also active. 

There is negligible opportunity for slip transfer from slip system 10 in Grain A to either of the highly active systems in grain B ($m' = 0.42$ for slip system 4 and $m' = 0.19$ for slip system 8). 
In each grain, the two most favored SS are on different planes and in different directions, such that each will generate forest dislocation barriers on the other system.
With these four favored SS, the hardening rate resulting from double-slip is higher than that in the single slip case (Fig. \ref{FS-StressStrain}).

In the double-slip bicrystal with the partially transparent boundary, the shear strain at the center adjacent to the boundary showed a significant reduction in the shear strains near the boundary, as shown in Figs. \ref{DS-shear-FS}c) and d).  With enabling transparency on only the best aligned pair, which are the most active of the SS (systems 2 in grain A and 8 in grain B), there was also a large decrease in accumulated shear strain in these SS (from 0.016\% to 0.002\% for a far-field applied strain of 2\%. The reduction was even higher with a fully opaque boundary (Figs. \ref{DS-shear-FS}e and f).  Thus,  the effect of making only one pair of SS transparent still leads to significant reduction in the activity of all SS in the vicinity of the boundary.

\subsubsection{Comparison between different GB slip criteria}

\begin{figure} 
 \centering
 \includegraphics[width=0.9\linewidth]{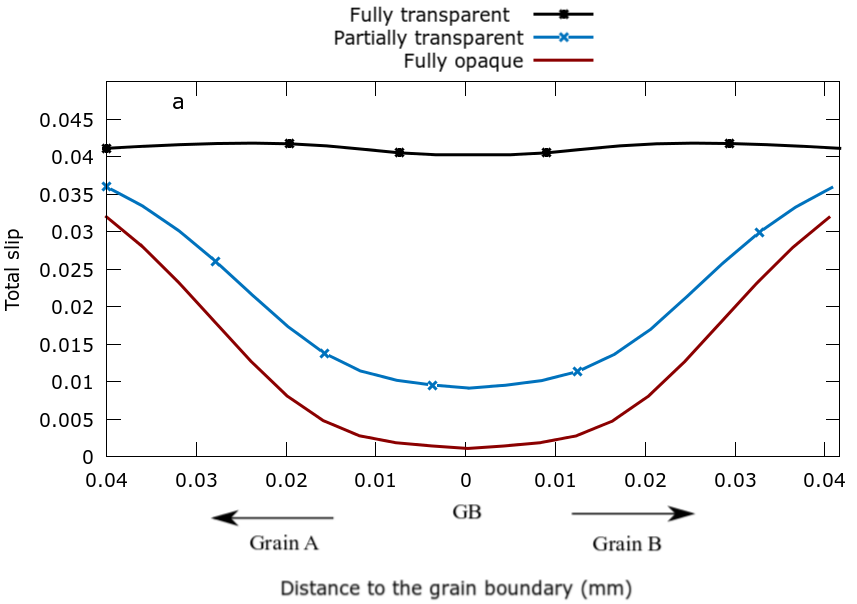}
  \includegraphics[width=0.9\linewidth]{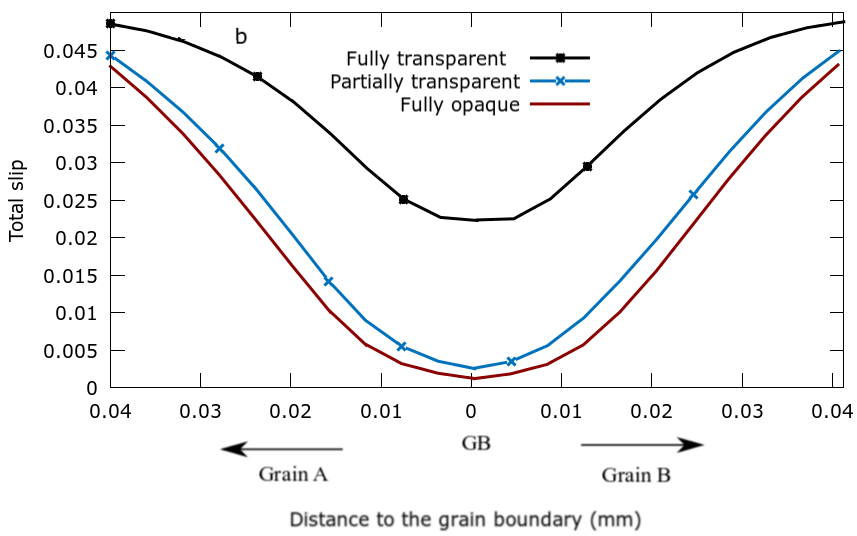}
\caption{Total slip accumulated on all the SS, $\Gamma$, along a line perpendicular to the GB in the center of the bicrystal for fully transparent, partially transparent and fully opaque GB. (a) Single slip  Al bicrystal. (b) Double slip Al bicrystal. 
The far-field applied strain was $2\%$ in all cases. Unconstrained boundary conditions.}
\label{SS-Total-Pstrain}
\end{figure}

The total slip $\Gamma$ ($= \sum_{\alpha} \int{ \mid \dot\gamma^{\alpha} \mid  dt}$) along a line perpendicular to the GB in the center of the bicrystal is shown in Fig. \ref{SS-Total-Pstrain}a for the single-slip Al bicrystal with fully transparent, partially transparent and fully opaque GB. The far-field applied strain was 2\% in all cases. In the case of the fully-transparent boundary, the total slip remain constant near the boundary, indicating that it was not an obstacle for slip transfer. Nevertheless, there is a noticeable hardening effect at the boundary (indicated by the reduction in the total slip) when only the most favored slip system is allowed to be transparent. This hardening is stronger  in the fully opaque case.  

The equivalent plot of the total slip for the double-slip bicrystal is shown in Fig. \ref{SS-Total-Pstrain}b. It is interesting to notice that there is a reduction in the total slip near the GB even in the case of the fully-transparent boundary. This hardening of the GB comes about because slip transfer is very easy between slip  system 8 in grain B with slip system 2 in grain A ($m'=0.98$) but not between slip system 10 in grain A and slip system 4 in grain B ($m'=0.42$). Thus, out of the two SS active in each grain, only one is favorably oriented for easy slip transfer. The hardening of the GB is more marked when slip transfer is only allowed for the pairs of SS with high $m'$ and, in fact, is very similar to the one found for the fully-opaque GB. 

\begin{figure} 
 \centering
 \includegraphics[width=\linewidth]{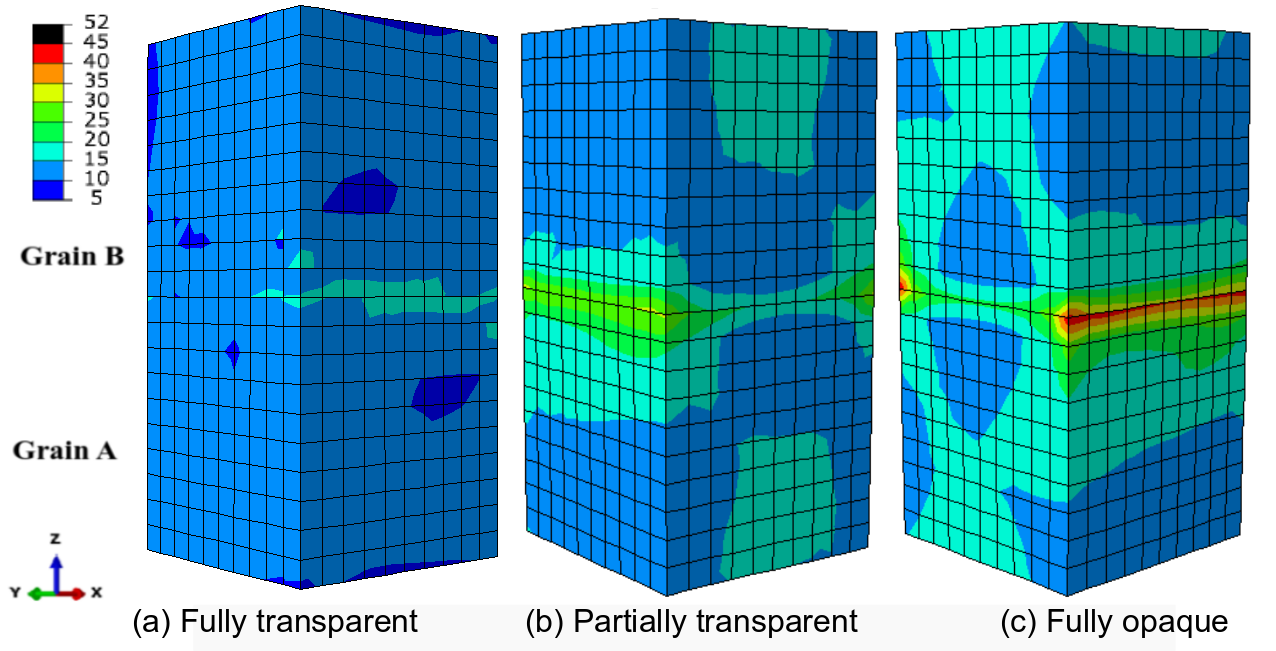}
\caption{Contour plots of the Von Mises stress for a single-slip  Al bicrystal with different GB properties. (a) Fully transparent, (b) Partially transparent, (c) Fully opaque.
The far-field applied strain was $2\%$ in all cases, under unconstrained boundary conditions. Stresses are expressed in MPa.}\label{SS-Vmises}
\end{figure}

\begin{figure} 
 \centering
 \includegraphics[width=\linewidth]{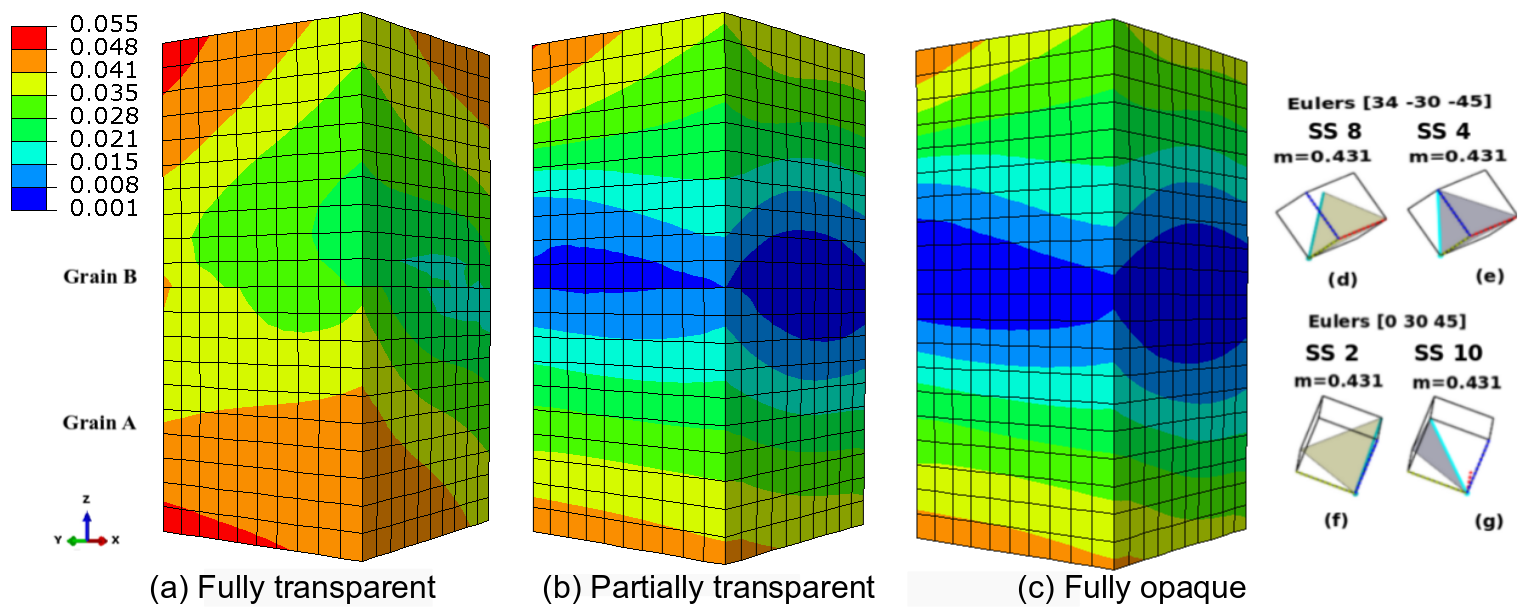}
\caption{Contour plots of the accumulated plastic slip on all the SS, $\Gamma$, for a double-slip Al bicrystal with different GB properties. (a) Fully transparent, (b) Partially transparent, (c) Fully opaque. The far-field applied strain was $2\%$ in all cases, under unconstrained boundary conditions. The unit cells show the crystal orientation of grains A and B and the two most highly activate SS. }\label{DS-Pstrain}
\end{figure}

\begin{figure} 
 \centering
 \includegraphics[width=\linewidth]{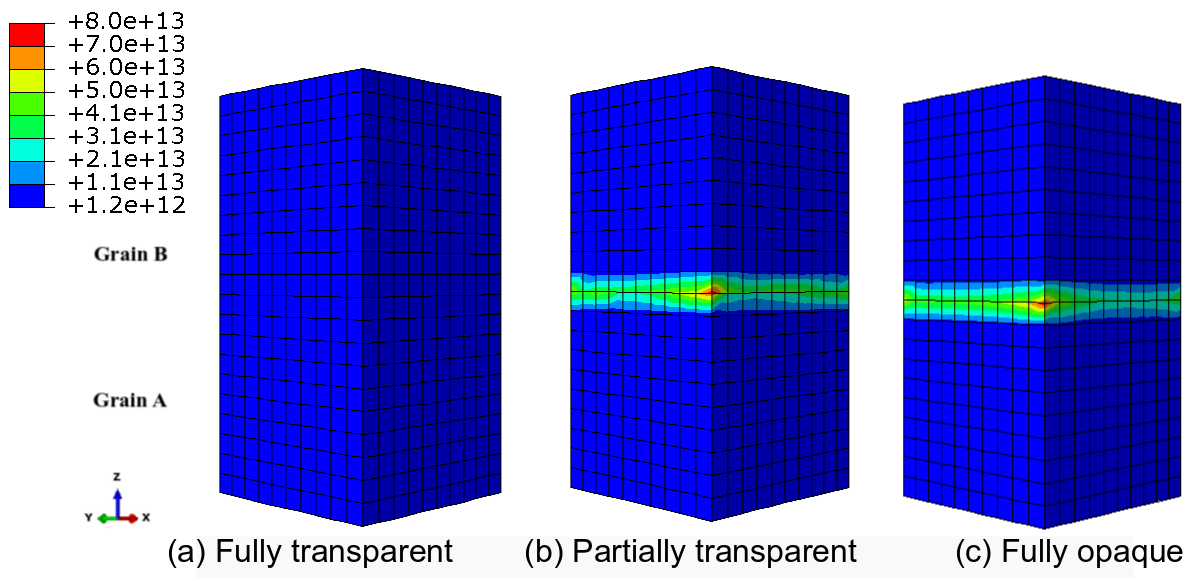}
\caption{Contour plots of the total dislocation density in all the SSor a double-slip Al bicrystal  with different GB properties. (a) Fully transparent, (b) Partially transparent, (c) Fully opaque. The far-field applied strain was $2\%$ in all cases, under unconstrained boundary conditions.The dislocation density is expressed in m$^{-2}$.}\label{DS-Dislocation}
\end{figure}

\begin{figure} 
 \centering
 \includegraphics[width=\linewidth]{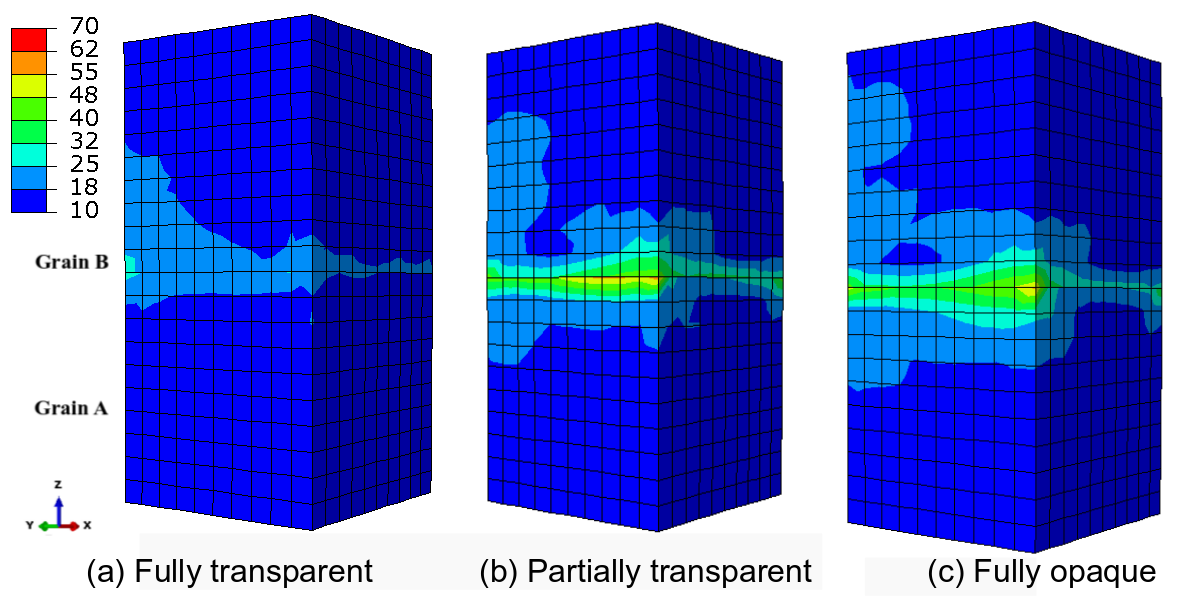}
\caption{Contour plots of the Von Mises stress for a double-slip Al bicrystal with different GB properties. (a) Fully transparent, (b) Partially transparent, (c) Fully opaque.
The far-field applied strain was $2\%$ in all cases, under unconstrained boundary conditions. Stresses are expressed in MPa.}\label{DS-Vmises}
\end{figure}

The influence of slip transfer at GBs on the deformation pattern can be assessed from Figs. \ref{SS-Pstrain}-\ref{SS-Vmises} for the single-slip bicrystal, and \ref{DS-Pstrain}-\ref{DS-Vmises} for the double slip bicrystal. These figures show the contour plots of the accumulated plastic slip on all SS, the total dislocation density and the Von Mises stress, for a single- and double-slip Al bicrystals, for the different boundary characteristics, namely fully transparent, partially transparent and fully opaque.

\begin{figure} 
 \centering
 \includegraphics[width=\linewidth]{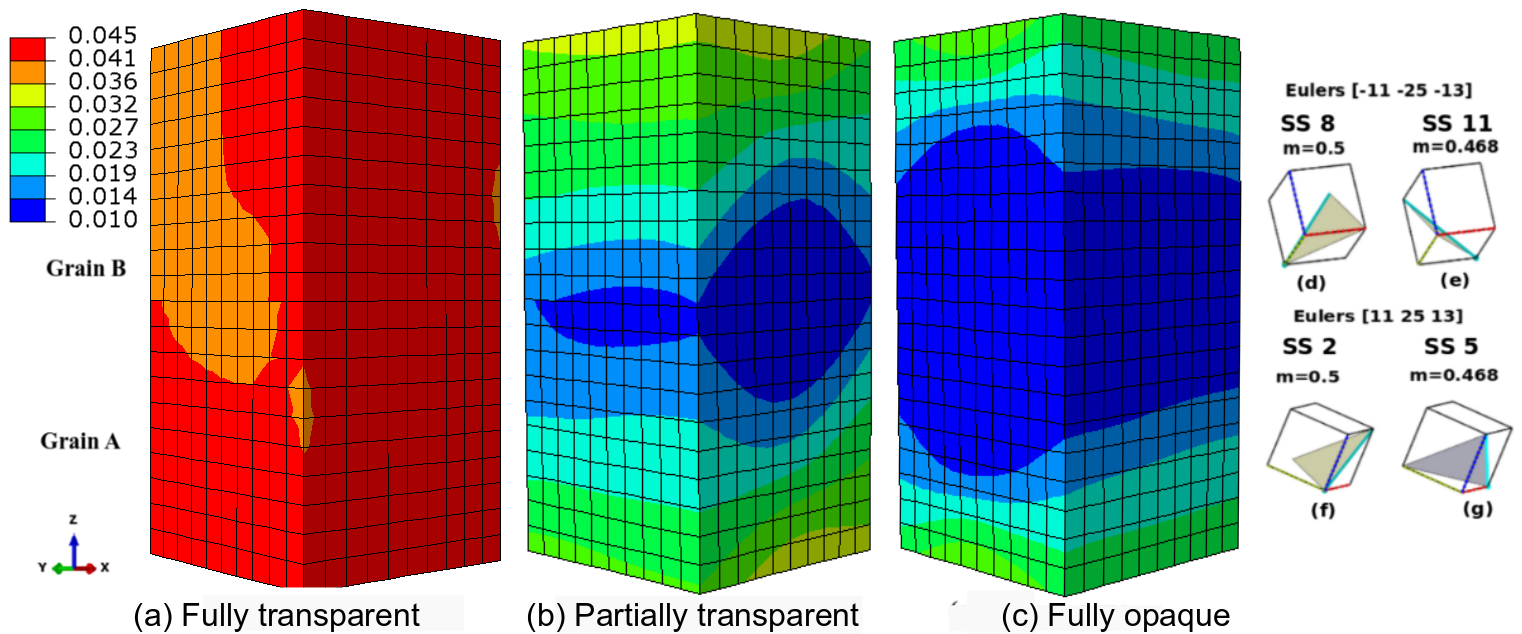}
\caption{Contour plots of the accumulated plastic slip on all the SS, $\Gamma$, for a single-slip Al bicrystal with different GB properties. (a) Fully transparent, (b) Partially transparent, (c) Fully opaque. The far-field applied strain was $2\%$ in all cases, under unconstrained boundary conditions.  The unit cells show the crystal orientation of grains A and B and the two most highly activate SS. }\label{SS-Pstrain.png}
\end{figure} 

\begin{figure} 
 \centering
 \includegraphics[width=\linewidth]{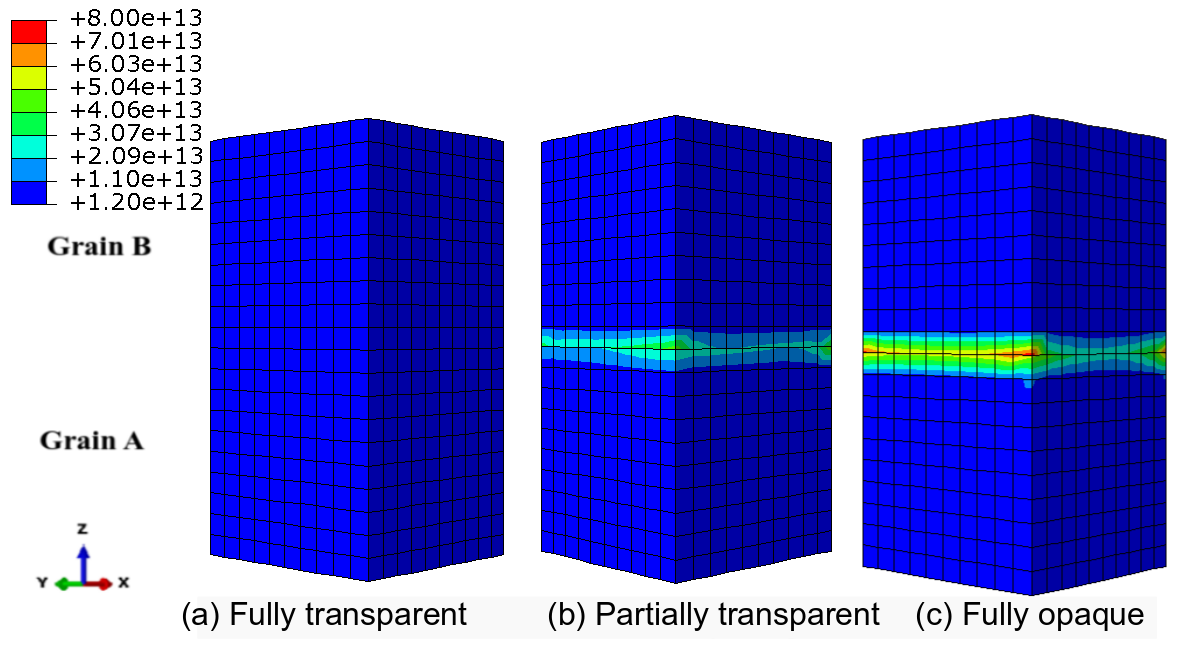}
\caption{Contour plots of the total dislocation density for all SS for a single-slip  Al bicrystal with different GB properties. (a) Fully transparent, (b) Partially transparent, (c) Fully opaque. The far-field applied strain was $2\%$ in all cases, under unconstrained boundary conditions. The dislocation density is expressed in m$^{-2}$.}\label{SS-Dislocation}
\end{figure} 

In the case of the fully transparent boundary, slip in one grain can progress into the next grain with apparent continuity for single and double slip, as shown in accumulated plastic slip contour plots in Figs. \ref{SS-Pstrain}a and \ref{DS-Pstrain}a. The continuity of the orange band through the boundary and of the red regions of maximum strain near the boundary illustrate that the boundary has no influence on the continuity of the dominant slip system in Fig. \ref{SS-Pstrain}a. 
Also, the Von Mises stress in Figs. \ref{SS-Vmises}a and the dislocation densities in Fig. \ref{SS-Dislocation}a, are fairly homogeneous throughout the microstructure for single slip. In the case of double slip, however, the total plastic slip near the GB is slightly lower than far-away (Fig. \ref{DS-Pstrain}a) because only one dominant slip system in each grain is suitable oriented for slip transfer, as indicated above. As a result, the contour plot of the Von Mises stress (Fig. \ref{DS-Vmises}a) shows as light stress concentration.

In contrast, there is more plastic deformation away from the boundary  than around the GB for the partially transparent and fully opaque boundaries, as shown in Figs. \ref{SS-Pstrain}b-c and  \ref{DS-Pstrain}b-c. The reduction in the plastic slip near the boundary is accompanied by an increase in the dislocation density due to the formation of dislocation pile-ups (Figs. \ref{SS-Dislocation}b-c and \ref{DS-Dislocation}b-c) and also by an increase in the Von Mises stress (Figs. \ref{SS-Vmises}b-c and \ref{DS-Vmises}b-c). In the case of opaque boundary, the dislocations are forced to accumulate at the interface, with no possibility to generate any further plastic slip, leading to a pileup of dislocations at the impenetrable interface, (Figs. \ref{SS-Dislocation}c and \ref{DS-Dislocation}c), and to local stress concentrations (Figs. \ref{SS-Vmises}c and \ref{SS-Vmises}c).  With the partially transmissive boundary, the pileup is less pronounced (Figs. \ref{SS-Dislocation}a-b and \ref{DS-Dislocation}a-b) and the stress concentration is reduced (Figs. \ref{SS-Vmises}a-b and \ref{DS-Vmises}a-b). The von Mises streses shown in Fig. \ref{SS-Vmises} correlates closely with the dislocation densities in Fig. \ref{SS-Dislocation}.

It is interesting to notice that the transition from fully-transparent to partially-transparent to fully-opaque GBs also leads to a gradual transition in all the fields (plastic slip, dislocation density and von Mises stress) near the GB in the case of single slip (Figs. \ref{SS-Pstrain}, \ref{SS-Dislocation} and \ref{SS-Vmises}). One important consequence of this observation is that hindering the slip transfer of the less active SS (partially-transparent interface) cannot be neglected although most of the plastic slip in both crystals is concentrated in two SS suitable oriented for slip transfer. If it is assumed that the GB is fully transparent for all SS, the predictions of all the field variables will be very different from those obtained if slip transfer along less-active SS with low $m'$ is inhibited. 

In the case of the double slip bicrystal, the contour plots of the field variables (plastic slip, dislocation density and von Mises stress) near the GB are very similar for the cases of partially-transparent and opaque boundary) and very different from those found for the fully-transparent boundary (Figs. \ref{DS-Pstrain}, \ref{DS-Dislocation} and \ref{DS-Vmises}). Thus, the more realistic simulations (partially-transparent GB) in the case of double slip bicrystal are closer to the fully-opaque GB, while they were in between the fully-transparent and the fully-opaque case for single slip bicrystals. This behavior can be attributed to the fact that slip transfer is always hindered in one slip system with high SF in each grain and that hindering the slip transfer of these active SS has major consequences in the accumulation of dislocations and in the hardening of the GB.

\subsection{Constrained periodic boundary conditions}

\begin{figure} 
\centering
\begin{subfigure}{.5\textwidth}
  \centering
  \includegraphics[width=\linewidth]{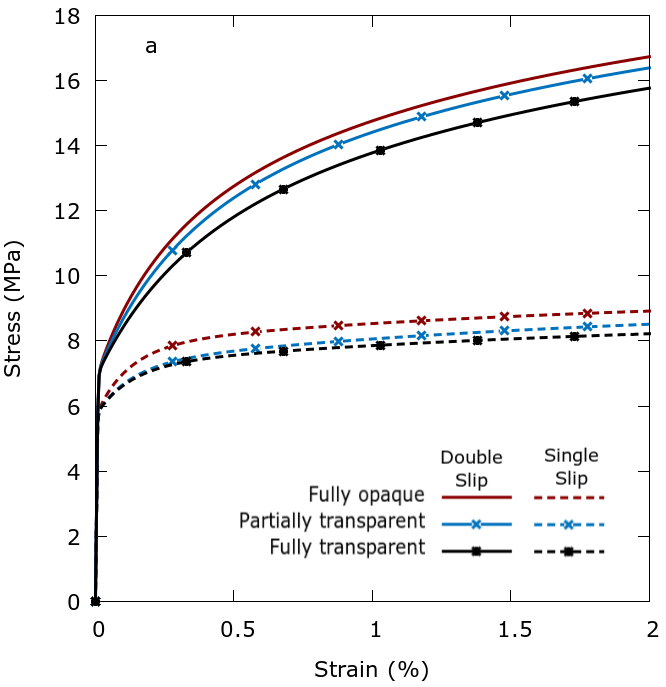}
  \caption{ }
  \label{fig:sub1}
\end{subfigure}%
\begin{subfigure}{.5\textwidth}
  \centering
  \includegraphics[width=\linewidth]{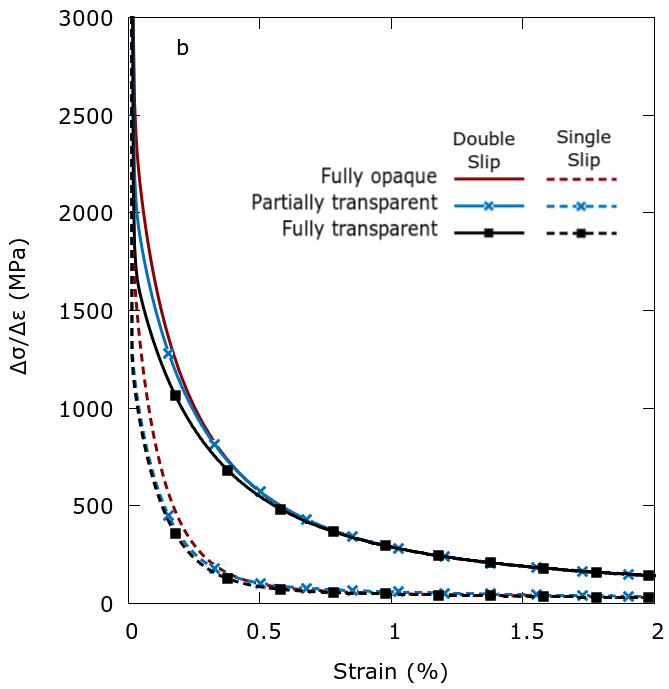}
  \caption{ }
  \label{fig:sub2}
\end{subfigure}
\caption{(a) Stress-strain curves and (b) strain hardening rate-strain curves under constrained perdiodic boundary conditions for the single-slip (dashed line) and double-slip (solid line) Al bicrystals as a function of the GB characteristics.  }
\label{PBC-StressStrain}
\end{figure}

The tensile deformation of single- and double-slip Al bicrystals were simulated under constrained periodic boundary conditions, and the corresponding stress-strain curves and hardening curves are plotted in Fig. \ref{PBC-StressStrain}. As with the unconstrained boundary conditions, the stress and strain hardening curves corresponding to partially-transparent boundaries are in between those of the fully transparent and fully opaque boundaries. It should be noted that the differences in the flow stress for different GBs were higher under constrained boundary conditions than in the case of unconstrained deformation. The magnitude of the flow stress in the case of single slip was similar for both boundary conditions but it was  40\% higher in the case of constrained deformation for bicrystals oriented for double slip (Fig. \ref{FS-StressStrain}a).

\begin{figure}  
 \centering
 \includegraphics[width=\linewidth]{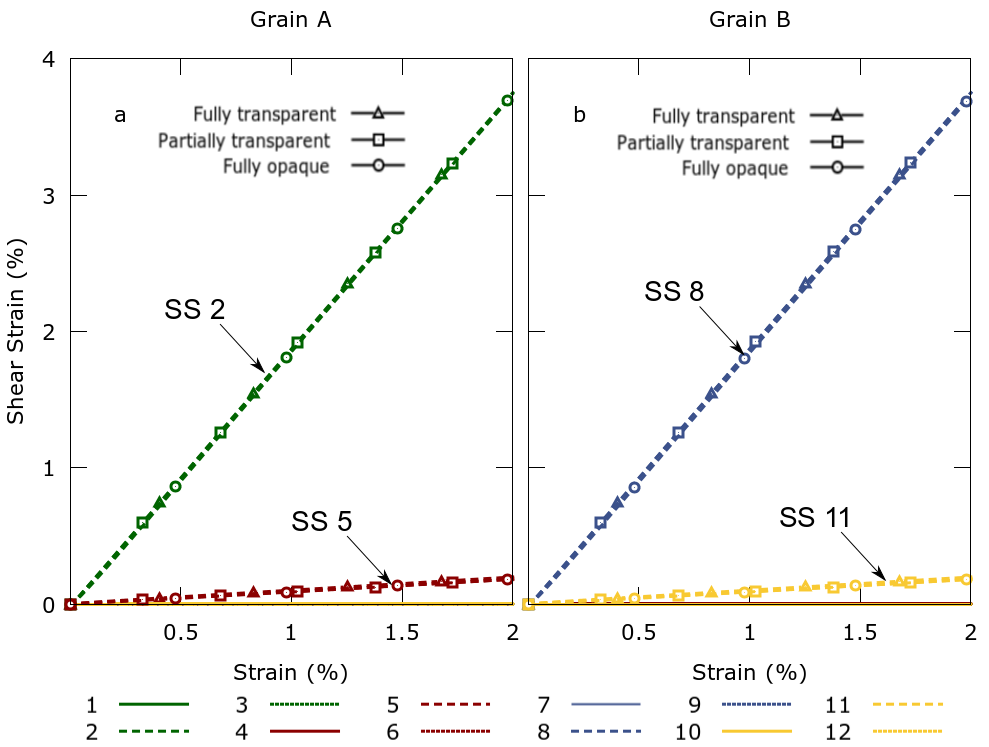}
\caption{Accumulated shear strain in the active SS at the center of the GBs for a single-slip Al bicrystal as a function of the far-field applied strain . The results for different types of GBs are indicated with different symbols. The far-field applied strain was $2\%$ in all cases under constrained periodic boundary conditions.}
\label{SS-shear-PBC}
\end{figure}

\begin{figure}  
 \centering
 \includegraphics[width=\linewidth]{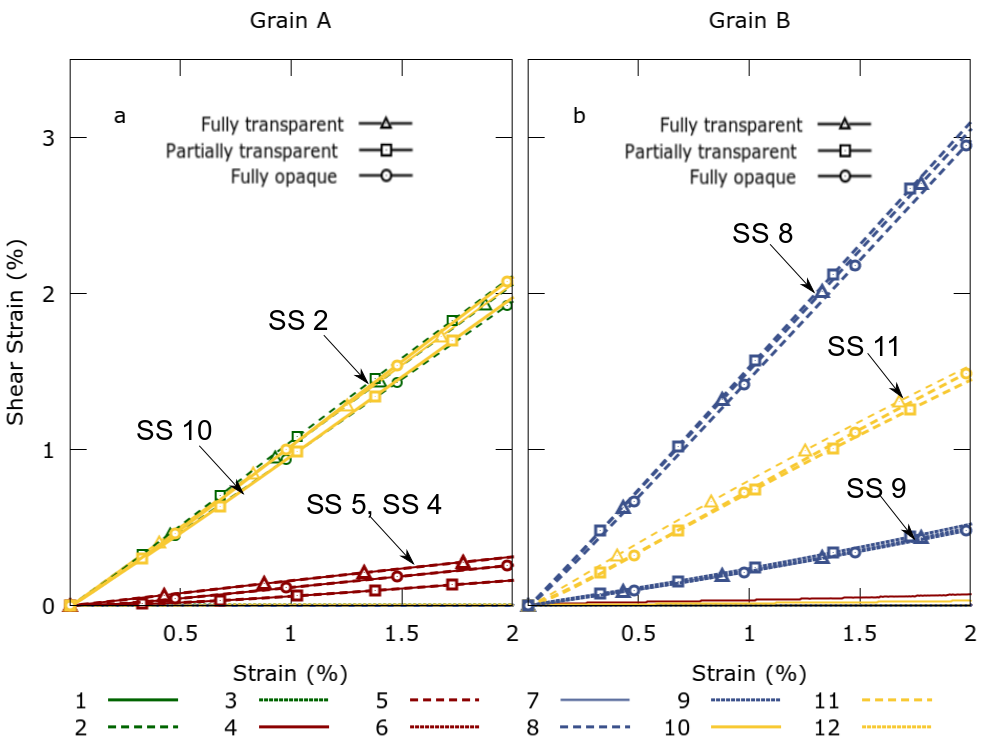}
\caption{Accumulated shear strain in the active SS at the center of the GBs for a double-slip Al bicrystal as a function of the far-field applied strain . The results for different types of GBs are indicated with different symbols. The far-field applied strain was $2\%$ in all cases under constrained periodic boundary conditions.
}\label{DS-shear-PBC}
\end{figure}

The accumulated shear strains in the active SS near the boundary are shown for both grains in Fig. \ref{SS-shear-PBC} for a single-slip bicrystal and in Fig. \ref{DS-shear-PBC} for a double-slip bicrystal, for the three different types of GBs, namely fully-transparent, partially-transparent and fully-opaque. Due to the constrained boundary conditions,  the differences in the shear strains in the active SS  among the three types of GBs are practically indistinguishable in  bicrystals oriented for single and double slip. Periodic boundary conditions do not allow strain gradients to develop along any surface, i.e., the desired overall strain gradient is imposed uniformly along the surface of the bicrystal and, consequently, every finite element in the simulation experiences the same strain path.

For the single-slip Al bicrystal subjected to periodic boundary conditions, the most favored SS with the highest SFs are activated (slip system 2 in crystal A and slip system 8 in crystal B) as shown in Fig. \ref{SS-shear-PBC}.  The shear strains on these dominant SS are about the same as the ones obtained under unconstrained boundary conditions.
The strains on the secondary slip system (slip system 5 in crystal A and slip system 11 in crystal B) are about 5\% of the strain on the primary slip system for all three boundary types, whereas the secondary slip system in the unconstrained model consistently carries about 8\% of the shear of the dominant system for the transparent boundary.   

In contrast to the single-slip bicrystal, the double-slip bicrystal deformed under constrained boundary conditions shows very different distributions of shear strain on the favored SS (Fig. \ref{DS-shear-PBC}) as compared to the bicrystal deformed unconstrained conditions (Fig. \ref{DS-shear-FS}). Under constrained deformation, the two most favored SS in grain A show nearly the same amount of shear strain, with a lesser amount of strain in the third highest SF slip system 5.  In grain B, the relative ordering of observed slip activity is quite different; the most strain occurs on slip system 8, followed by 11,  9, and a little on 4 and 10, whereas the order is 8, 4, 11, 9 and 10 in the bicrystal deformed under unconstrained boundary conditions.

\begin{figure}  
 \centering
 \includegraphics[width=\linewidth]{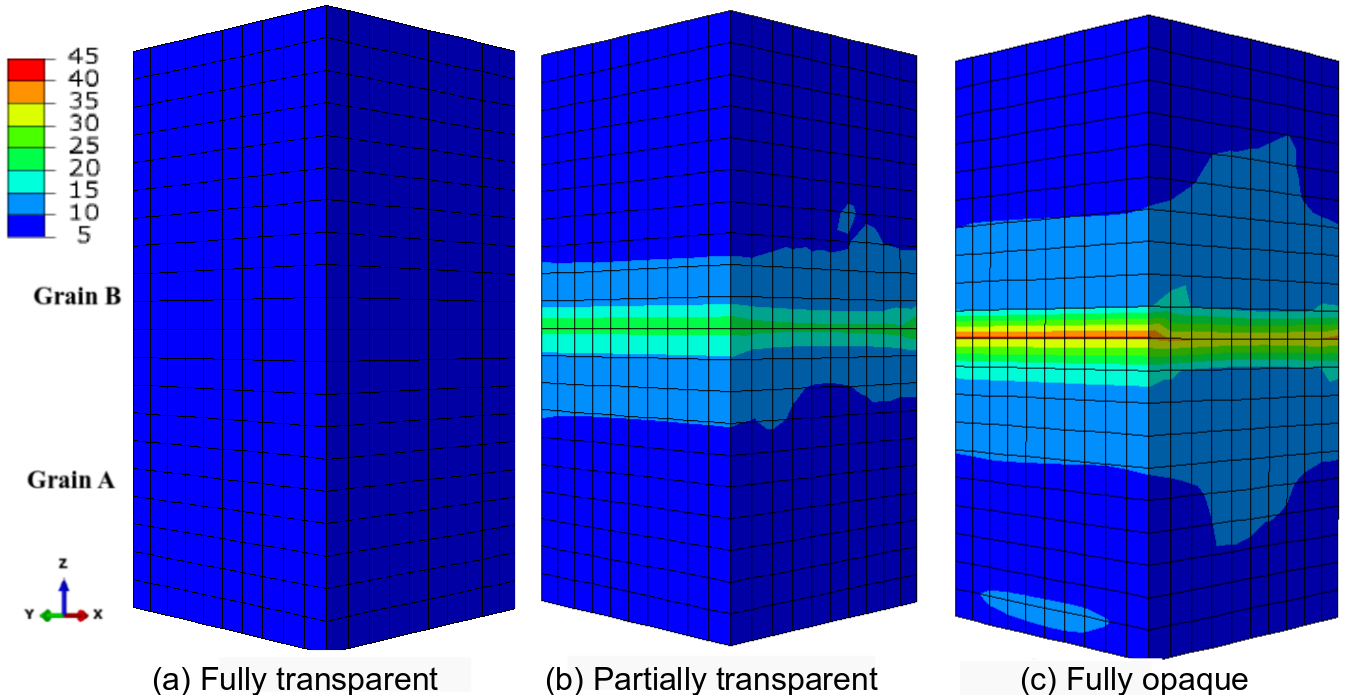}
\caption{Contour plots of the Von Mises stress for a single-slip Al bicrystal with different GB properties. (a) Fully transparent, (b) Partially transparent, (c) Fully opaque.
The far-field applied strain was $2\%$ in all cases, under constrained boundary conditions. Stresses are expressed in MPa.}\label{SS-PBC-Vmises}
\end{figure}

\begin{figure}  
 \centering
 \includegraphics[width=\linewidth]{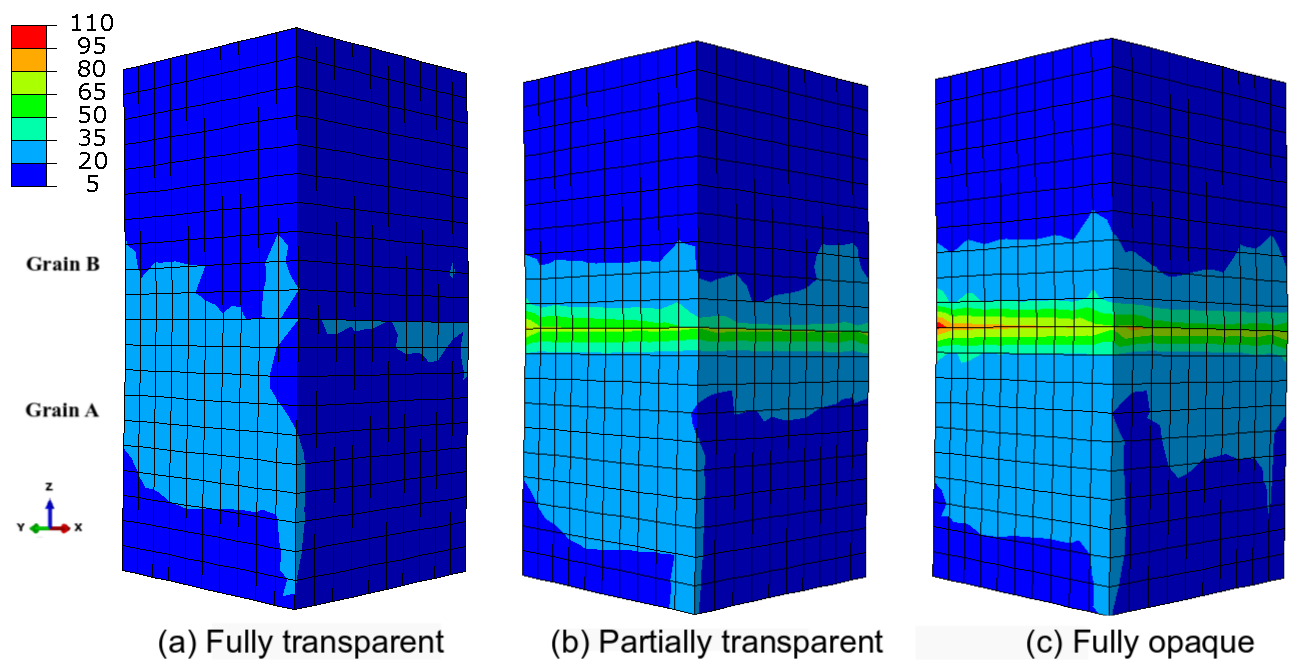}
\caption{Contour plots of the Von Mises stress for a double-slip Al bicrystal with different GB properties. (a) Fully transparent, (b) Partially transparent, (c) Fully opaque.
The far-field applied strain was $2\%$ in all cases, under constrained boundary conditions. Stresses are expressed in MPa.}\label{DS-PBC-Vmises}
\end{figure}

Regardless of these differences in the active systems for the bicrystal oriented for double slip for unconstrained and constrained boundary conditions, the effect of the GB characteristics on the field variables did not change. The contour plots of the total slip accumulated on all the SS is not plotted because there is no spatial variation. The contour plots of the Von Mises stress for single- and double-slip Al bicrystals are shown in Figs. \ref{SS-PBC-Vmises} and \ref{DS-PBC-Vmises}, respectively, for the different GBs.  The Von Mises stress is nearly homogeneous throughout the sample, Figs. \ref{SS-PBC-Vmises}a) and \ref{DS-PBC-Vmises}a) for the transparent GB, while a stress concentration develops near the GB due to a pileup of dislocations at the impenetrable interface  for the opaque GB (Figs. \ref{SS-PBC-Vmises}c and \ref{DS-PBC-Vmises}c).  The stress concentration depends on whether the bicrystal deforms by single or double slip despite the fact that the deformation imposed by the surfaces was homogeneous.  As in the unconstrained boundary conditions, the partially-transparent boundary is closer to the transparent case for the single-slip bicrystal, and closer to the opaque case for the double-slip bicrystal (Figs. \ref{SS-PBC-Vmises}b and \ref{DS-PBC-Vmises}b).

\subsection{Experimental GB}

The $m'_{\alpha\beta}$ values for all pairs of SS in the experimental GB are provided in Table \ref{slipsystems}. The consequence of the small 13.4$^\circ$ misorientation leading to all systems having very high ($>$ 0.95) $m'_{\alpha\beta}$  values with the corresponding SS in the neighbouring grain is evident, and these $m'_{\alpha\beta}$ values are in bold font in Table \ref{slipsystems}. 
There are eight pairs of SS (10, 1, 6, 3, 12, 9, 4 and 7) with $m'_{\alpha\beta} \ge $ 0.95 and with high SF ($\ge$ 0.30) for both SS across the GB. 
The experimental results indicate that this GB is transparent
from the viewpoint of slip transfer, where at least one SS shows well correlated slip traces.  From the observations in \cite{BAP19}, a threshold of 0.8 was observed for slip transfer based upon the product of $m'_{\alpha\beta}$ times the sum of the SFs for each slip system pair.  Table \ref{times} provides these products for the 8 SS that have high SFs, indicating that four of the SS have a value larger than 0.8. Thus,  these four pairs of SS (9-9, 12-12, 1-1, 10-10) were made transparent and the rest opaque in the simulations for the partially transparent GB shown below.  This provides an exemplary case to compare simulations with the experimental observations.
The experimental results indicate that this GB is very transparent from the viewpoint of slip transfer and it seems an exemplary case to assess the slip transfer criterion based on a threshold value for $m'_{\alpha\beta}$.

\begin{table}
\caption {Product of  $m'_{\alpha\beta}$ times the sum of the corresponding SFs ($SF_{\alpha} + SF_{\beta}$) for pairs of SSs in the experimental GB in Fig. \ref{exp}. Only the SSs with high SF are included in the table.}\label{times}
\centering
   \begin{tabular}{ l c c c  }
   \hline
   SS pair  & $SF_{\alpha} + SF_{\beta}$ & $(SF_{\alpha} + SF_{\beta}) m'_{\alpha\beta}$  \\ 
  9-9  &  0.86 &  0.834 \\ 
  12-12  & 0.86  &  0.832 \\ 
  1-1  & 0.86 & 0.826  \\ 
  10-10  &  0.85 & 0.818  \\ 
  6-6  &  0.71 & 0.689  \\ 
  3-3  &  0.71 & 0.685  \\ 
  7-7  &  0.70 & 0.665  \\ 
  4-4  &  0.68 &  0.647  \\   
          \hline
   \end{tabular}
 \end{table}

In the experimental evidence in Fig. \ref{exp}b) and c), the SS 1 (green) is the most noticeably active in both grains, but its SF is ranked  3rd in grain A and 2nd in grain B (it is possible that SS 3 also contributed to these contiguous slip traces, but its activity would be less visible and it has a lower SF in both grains).  The SS 9 (blue, or possibly system 7, but it has a lower SF) is also apparent in both grains, with weak topography associated with a nearly invisible slip direction, but its presence is stronger in grain B.  There is no evidence of slip transfer on this slip plane, as the traces fade out close to the boundary. 

\begin{figure} 
 \centering
 \includegraphics[width=\linewidth]{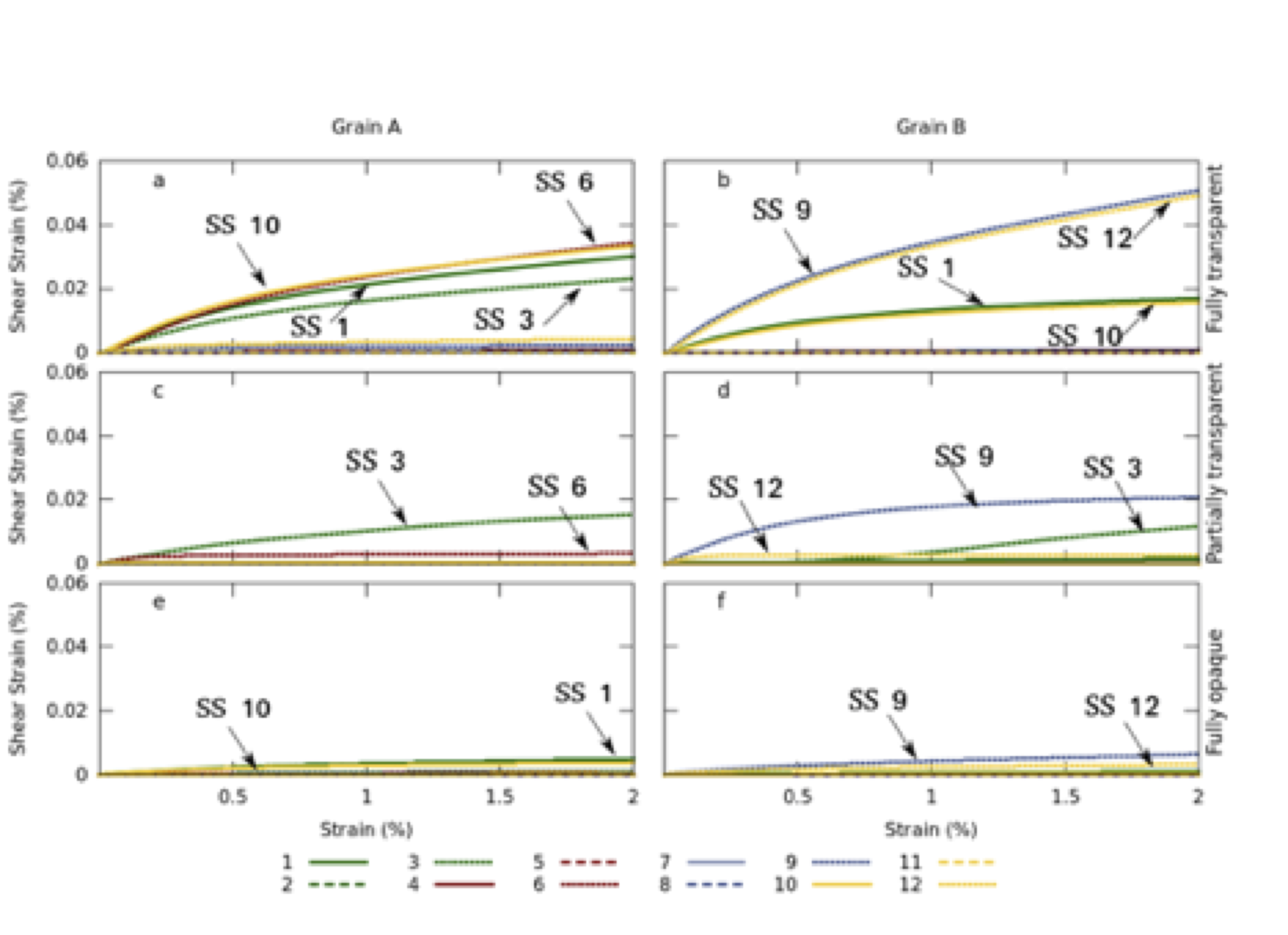}
\caption{Evolution of the shear strain accumulated on each slip system is plotted for each crystal very close to the GB at points A and B as a function of the applied strain for the experimental GB. The results for fully-transparent, partially-transparent and fully-opaque GBs are included. }
\label{Shear-exp}
\end{figure}

The simulated shear strain accumulated on each slip system at points A and B
(very close to the GB in the center of the sample in each crystal in Fig. \ref{RVE}a)
are plotted in Fig. \ref{Shear-exp} as a function of the applied strain for the three different GB models.  It is evident that the amount of strain near the boundary increases with increasing transparency.  The simulations with the transparent GB show some degree of agreement with the experiment in that the experimentally observed SS 1 is among the top four active systems in both grains, though SS 10 shows a greater amount of activity in both grains, along with less visible systems (dotted lines).  Interestingly, the low visibility SS 9 is very active in grain B but not in grain A, consistent with the experiment.  For the partially transparent boundary, the relative activity of the SS is quite different, and SS 1 is not appreciably active, though SS 9 is very active in grain B (opposite of experimental observations).  The SS 3 is very active in both grains, which has the same trace as SS 1 that is apparently dominant in the experiment.  This could be consistent with the experiment if the observed slip traces are actually from a less visible slip direction.  For the opaque boundary, the SS 1 is the most active in grain A, and the SS 9 is the most active in grain B.  This too shows a reasonable connection with the experiment, but it is not any more convincing than either of the other two cases.  

Considering the relative ranking of SS based upon uniaxial tension assumptions, the relative ranking from the simulations differ significantly, especially for grain B, where they are in nearly the reverse order; the most active slip system in all three simulations has the 6th highest SF.  The relative ranking of activity of SS in grain B is more consistent for the three simulations than in grain A, which has a variety of relative rankings depending on the simulation.

While the agreement is not perfect, the simulation has significantly different boundary conditions than the simulation, as the simulation has free surfaces on all four sides, and the experiment has one free surface from which observations come.  The simulation tracks slip activity in the center of the sample.  Given these rather important differences, it is reasonable that the agreement is not perfect.  Nevertheless, the degree of agreement between the experiment and simulation is at least semi-quantitatively credible, suggesting that the simulation method is able to capture a reasonable approximation of physically meaningful slip behavior as influenced by the grain boundary.  This comparison is not a validation of the model, but it points toward consistency.

\section{Conclusions}

The effect of slip transfer on the deformation mechanisms of Al bicrystals deformed in tension was explored using a rate-dependent dislocation-based crystal plasticity model. The critical resolved shear stress on each slip system depends the dislocation density through the Taylor model and the evolution of the dislocation density in each slip system follows a Kocks-Mecking law. In the latter, the net storage rate of dislocations per slip system is the sum of a positive storage rate governed by a length scale which was the lowest of the dislocation mean free path or distance to the GB, and a negative term accounting for dynamic recovery. This formulation naturally leads to the formation of pile-ups at the GBs and, thus, to fully-opaque (blocking) GBs \citep{HSL18}. If the GB term is not included in the Kocks-Mecking law, the GBs are fully-transparent from the viewpoint of slip transfer.  In addition, partially-transparent GBs were introduced in the model through the use of the Luster-Morris parameter. When the Luster-Morris parameter between two slips systems in neighbouring grains is above a critical value (indicating that both systems have highly geometrically aligned SS), the  GB term in the Kocks-Mecking law is not included for these SS, allowing the slip transfer between them.

Two ideal bicrystals were oriented for preferred slip on a single slip system and for slip on two SS with the same high Schmid factor, and simulations were carried out under unconstrained and constrained periodic boundary conditions. The results of the numerical simulations showed that the plastic deformation of the bicrystal by dislocation glide occurs along the SS that have high SF. The stress-strain curves for double-slip bicrystals exhibit higher hardening than those for a single- slip bicrystal because of latent hardening. 

Modifications of the GB character led to important changes in the deformation mechanisms at the GB. In general, bicrystals with fully-opaque boundaries deformed under unconstrained boundary conditions showed an increase in the dislocation density near the GB, which was associated with an increase in the Von Mises stress. Moreover,  the plastic strains at the boundary were smaller than in the bulk. In contrast, the dislocation pile-ups and the stress concentration were less pronounced in the case of partially-transparent boundaries as the slip in one grain can progress into the next grain with some degree of continuity. No stress concentrations were found at the GBs for fully-transparent boundaries, and there was continuity of strain across the boundary, which is not typical of most experimentally observed GBs \citep{HNP18, BAP19}. For constrained periodic boundary conditions, the applied deformation is imposed and the accumulated plastic slip is almost the same for different types of GBs. The influence of GB nature (either opaque, partially-transparent or transparent) leads to higher local stresses to accommodate the imposed uniform deformation in the boundary as the opacity of the GB increases.

Nevertheless, the magnitudes of the dislocation pile-ups and stress concentrations at the GBs depended on the details of crystallographic slip and boundary conditions and this was demonstrated by comparing the idealized bicrystals oriented for single and double slip. The single slip oriented bicrystal enables easy slip transmission through the boundary, making the boundary more similar to a transparent condition. In contrast, when two SS have similar potential for activation in both grains, but only one pair is aligned, the strain is higher on the slip system pair that is aligned, but the effects of latent hardening cause the boundary to become almost as opaque as a fully opaque boundary.  Therefore the conditions for slip transfer depend sensitively on the number of active SS in operation in the neighborhood.  This outcome suggests that most boundaries will lead to nearly opaque conditions, and that some boundaries will be partially transparent.  Thus. incorporating slip transfer into a crystal plasticity model that assumes that GBs are fully-opaque \citep{HSL18}  will lead to a slightly softer flow behavior, but whether this can better predict the local distribution of strain effectively remains to be assessed.

Finally, the model was applied to a particular experimentally observed
GB in which slip transfer was clearly operating. The appropriate application of the model to the most favored SS predicted significantly different levels of slip system activity than either the transparent or opaque cases, but in all three cases, the experimentally observed SS were among those that showed high activity in the simulations. 
  Nevertheless, the actual boundary conditions in the experiment and the simulation are not the same, so this represents a reasonable demonstration that the model can produce a meaningful outcome, but it is not a validation of the model.

\section{Acknowledgements}
This investigation was supported by the European Research Council (ERC) under the European Union's Horizon 2020 research and innovation programme (Advanced Grant VIRMETAL, grant agreement No. 669141) and by the HexaGB project supported by the Spanish Ministry of Science (reference RTI2018-098245). T.R. Bieler acknowledges the support from the Talent Attraction program of the Comunidad de Madrid (reference 2016-T3/IND-1600) for his sabbatical in Madrid and R. Alizadeh also acknowledges the support from the Spanish Ministry of Science through the Juan de la Cierva program (FJCI-2016-29660).  TRB also acknowledges support from the Department of Energy Office of Basic Science via grant DE-FG02-09ER46637.



\end{document}